\documentclass[aps,pre,reprint,superscriptaddress,showpacs,longbibliography]{revtex4-1}
\usepackage{amsmath,amssymb,graphicx,units}
\usepackage[usenames,dvipsnames]{color}
\usepackage[pdftex, colorlinks=true, linkcolor=blue, citecolor=blue, urlcolor=blue]{hyperref}
\usepackage[toc,page]{appendix}

\renewcommand{\vec}[1]{\boldsymbol{#1}}

\newcommand{\ipcms}{Universit{\'e} de Strasbourg, CNRS, Institut de Physique et Chimie des Mat{\'e}riaux de Strasbourg, UMR 7504, F-67000 Strasbourg, France}
\newcommand{\hqs}{HQS Quantum Simulations, Haid-und-Neu-Str.~7, 76131 Karlsruhe, Germany}

\begin{document}
\title{Energy Stability of Branching in the Scanning Gate Response of Two-Dimensional Electron Gases with Smooth Disorder}

\author{Keith R.\ Fratus}
\affiliation{\ipcms}
\affiliation{\hqs}
\author{Rodolfo A.\ Jalabert}
\affiliation{\ipcms}
\author{Dietmar Weinmann}
\affiliation{\ipcms}

\begin{abstract}
The branched pattern typically observed through the scanning gate microscopy (SGM) of two dimensional electron gases in the presence of weak, smooth disorder has recently been found to be robust against a very large shift in the Fermi energy of the electron gas. We propose a toy model, where the potential landscape reduces to a single localized feature, that makes it possible to recast the understanding of branch formation through the effect of caustics in an appropriate set of classical trajectories, and it is simple enough to allow for a quantitative analysis of the energy and spatial dependence of the branches. We find the energy stability to be extremely generic, as it rests only upon the assumptions of weak disorder, weak scattering, and the proportionality of the SGM response to the density of classical electron trajectories. Therefore, the robustness against changes of the electron's Fermi energy remains when adopting progressively realistic models of smooth disorder.\\[3mm]
Journal reference: \href{https://doi.org/10.1103/PhysRevB.100.155435}{Phys. Rev. B 100, 155435 (2019)} 
\end{abstract}

\maketitle

\section{Introduction}

Over the last two decades, developments in the field of scanning gate microscopy (SGM) have allowed for the detailed investigation of nanostructured devices based on a two-dimensional electron gas (2DEG) where different confining geometries can be defined; among them, a quantum point contact (QPC) \cite{topinka2000imaging,topinka2001coherent,topinka2003imaging,
leroy2005imaging,jura2007unexpected,jura09a,braem2018stable}, a
quantum billiard \cite{crook03b,polt16}, a ballistic ring \cite{martins07a,pala08a}, an electron wave-guide \cite{steinacher2016}, and a tunable electron cavity \cite{steinacher2018scanning}. The change in electrical conductance of the nanostructure under the effect of a charged atomic force microscope raster scanning above the surface of the device results in a mapping that provides a  characterization of the nature of the device well beyond that of a conventional electrical transport measurement \cite{sellier2011review}. This space-dependent data needs to be interpreted in order to extract useful information about the 2DEG and the nanostructure. 

Two limiting cases appear in an SGM setup. In the non-invasive regime, the voltage applied to the tip is weak enough as to result in a small perturbation of the transport problem. In the invasive regime, the tip strength is sufficiently strong as to create a divot at the level of the 2DEG under the tip. In this last case, the presence of a depletion disk much larger than the Fermi wavelength results in the backscattering of the impinging electrons. Recent experiments using tunable reflectors bridged the gap between these two limits by modulating the tip strength and the electronic confinement \cite{steinacher2018scanning}.

In the non-invasive regime, the SGM response is given by the unperturbed scattering wave-functions describing the nanostructure, and a relationship with the local density of states can be established in the case of a QPC operating very close to the condition of perfectly transmitting channels \cite{jalabert2010measured,gorini2013theory,ly2017partial}. In the invasive regime, the SGM map on a weakly-disordered 2DEG surrounding a QPC is not uniform through space, but is rather organized into thin, collimated structures, typically referred to as ``branches''  \cite{topinka2000imaging,topinka2001coherent,topinka2003imaging,kola}. Already the pioneering investigations of branching \cite{topinka2001coherent,topinka2003imaging} provided quantum and classical simulations of the electron flow in the 2DEG outside the QPC indicating that a classical approach is sufficient for the description of the branches. More complicated patterns are observed when the nanostructure is an electron wave-guide \cite{steinacher2016} or a ballistic cavity \cite{steinacher2018scanning}. 

The filamentary structure of the branches is a striking feature of the above-cited invasive SGM setup, since the electrons propagate along the 2DEG almost ballistically over the disorder potential. The electrons only suffer small-angle scattering because the disorder is rather smooth, with long-range spatial correlations, and weak enough as to have an amplitude significantly smaller than the Fermi energy. Interestingly, this branching phenomenon is not limited to 2DEG systems, and similar behavior has been observed in a variety of other physical phenomena, ranging from the propagation of ocean waves \cite{heller2008refraction,hegewisch2005ocean} and the focusing of tsunamis \cite{heller2016physics} to microwave transport experiments \cite{hohmann2010freak}, light in thin dielectric films \cite{patsyk2019light}, and electron flow in Dirac solids \cite{mattheakis2018emergence}.

Previous work in the case of disordered 2DEG systems has identified the local bumps and dips of the disorder potential as being responsible for forming caustics, or localized, singular concentrations of classical trajectories \cite{topinka2003imaging,topinka2001coherent} that result in the observed branching effect. The example of an incoming plane wave with parallel rays that are focused by the effect of a potential dip has been used to visualize the existence of directions exhibiting an increased density of scattered rays \cite{heller2003branching,vaniheller2003,maryenko2012branching}, as well as to quantify the statistics of branch formation in a weak correlated random potential \cite{kaplan2002statistics}.

The observed branches exhibit a surprising stability with respect to different parameters. In the case of ocean waves, once the branches are formed, the wave-front remains stable as it propagates over long distances, despite the dispersion generated by the potential fluctuations \cite{wolfson2001stability}. In the case of disordered 2DEG systems the structure of the branches was found to be stable at distances far away from the QPC when the latter is shifted laterally by the application of unequal potentials to its two gates \cite{jura2007unexpected}. Recently, changes in the electrons' Fermi energy have been seen to leave the shape and location of many of these branches largely intact, though often with a change in their relative intensity \cite{braem2018stable}.

In this work, we focus on explaining the recently observed stability of the branching pattern with respect to a relatively large shift in the Fermi energy that exceeds the amplitude of the fluctuations of the disorder potential, as observed in Ref.\ \cite{braem2018stable}. We introduce a toy model of disorder, describing the scattering of classical electron trajectories from a single localized feature of the disorder potential that accounts for the branch formation and the stability of the branches with respect to changes in the electron Fermi energy. In spite of the considerable simplicity of the above model, the identified mechanisms are still found at work when we consider more elaborate descriptions of the smooth disorder present in a 2DEG. The advantage of the toy model is to allow for a reformulation of the branching problem which is sufficient to understand the phenomenon of energy stability at the quantitative level, while avoiding any complications not directly related to this behavior. 

This paper is organized as follows. In Sec.\ \ref{sec:mech}, we present a toy model designed to represent a single localized feature in a disorder potential. Numerical and analytical results for this model indicate that it is capable of leading to branch formation which is stable with respect to a shift in energy. We examine the role that energy plays in the formation of these branches through its influence on the scattering of individual classical trajectories and explain how the qualitative features of this model lead to the energy stability of the branches. In Sec.\ \ref{sec:enStabGen} we argue why these features should be generic for any situation in which the scattering is sufficiently weak. In Sec.\ \ref{sec:mechDis}, we compare the results from the toy model with those generated using more realistic descriptions of disorder, to motivate that our simple model is sufficient to capture both the formation and energy stability of experimentally realistic branches. We conclude in Sec.\ \ref{sec:con}.

\section{Branch Formation and Stability in a Toy Model}\label{sec:mech}

\begin{figure}
\centering
\includegraphics[width=\columnwidth]{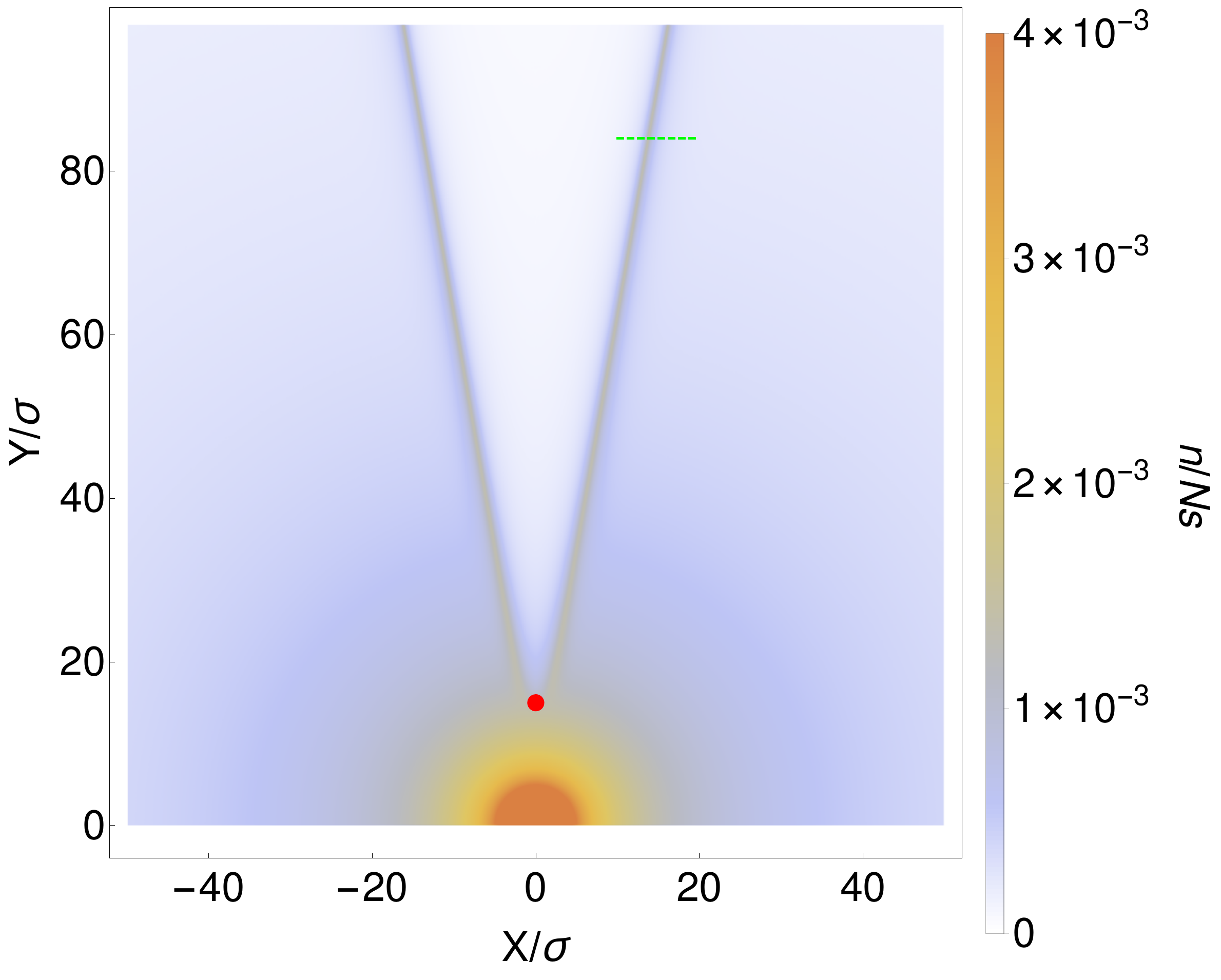} 
\caption{Density of classical electron trajectories as a result of scattering from a localized, repulsive Gaussian bump (indicated by the red dot) with width $\sigma$ and height $V_{0}$, illustrating the formation of well-collimated branches. The distance between the center of the bump and the QPC is taken to be $a = 15 \sigma$, with the QPC taken to be at the origin. The energy is tuned to be $E = 8.23V_{0}$. The numerical simulation involves $500,000$ trajectories distributed uniformly across an angular range of $\pi$ radians, for an angular trajectory density of $N \approx 159,154$ trajectories per radian. The density is computed by counting the number of trajectories $n$ which pass within a radius of $r = s / 2$ of a given lattice site on a grid of spacing $s = 0.02\sigma$. The green dashed line indicates the horizontal cut considered in Fig.\ \ref{fig:hCP}.}
\label{fig:zB}
\end{figure}
The formation of branches and their robustness with respect to changes of the electron energy, appearing in a smooth disorder potential, can already be understood in a classical toy model with a single localized feature in an otherwise clean potential. The detailed shape of this feature is not important; we simply require that a.) the feature consists of an isolated local minimum or maximum whose amplitude is less than the electron energy, and that b.) this local feature decays sufficiently ``quickly'' over some characteristic length scale. In the context of our model, we refer to the Fermi energy of the electrons as simply the ``energy,'' to emphasize its classical nature.

The classical trajectory-counting description, already employed in some of the first papers on the subject \cite{topinka2001coherent,topinka2003imaging}, identifies the SGM response with the density of classical trajectories at the position of the tip when the latter is not present. Such a description 
relies on two assumptions. Firstly, we treat the scattering off of the tip in the semi-classical limit, since the size of the divot is much larger than the Fermi wavelength of the electrons, while the electrostatic potential landscape presents small amplitude modulations and smooth spatial variations such that the changes of the electron momentum over a wavelength remain small when compared to the momentum itself \cite{merzbacher_quantum_2nd}.  
The semi-classical approach allows to show that, in the leading order in $\hbar$, the change in resistance by the effect of an invasive tip is given by the fraction of classical trajectories that hit the depletion divot and get back to the QPC \cite{jalabert2016}. 
This leading-order (incoherent) contribution stemming from the diagonal terms in the semiclassical expression of the conductance dominates over the off-diagonal (coherent) one, which in addition is suppressed by some amount of decoherence that can never be avoided \cite{metzger2013}.
Secondly, we surmise that among the trajectories hitting the tip almost head-on, the subset that make it back to the initial electrode represents a constant fraction of the total. That is, the probability of being reflected back through the QPC is independent of the position of the tip.
Detailed numerical calculations \citep{braem2018stable} have recently confirmed the approximate validity of the proportionality between the local trajectory density in the absence of the tip and the tip-induced conductance change in the case of a smooth disordered potential.  
In order to be consistent with the semi-classical approximation, the QPC is assimilated to a point source and we simply  assume that the angular distribution of the emerging classical trajectories is smooth.

To properly address the physics of scattering within the toy model, we treat the case of a local minimum and a local maximum separately.

\subsection{The case of a local maximum}\label{subsec:localmax}

\begin{figure}
\centering
\includegraphics[width=\columnwidth]{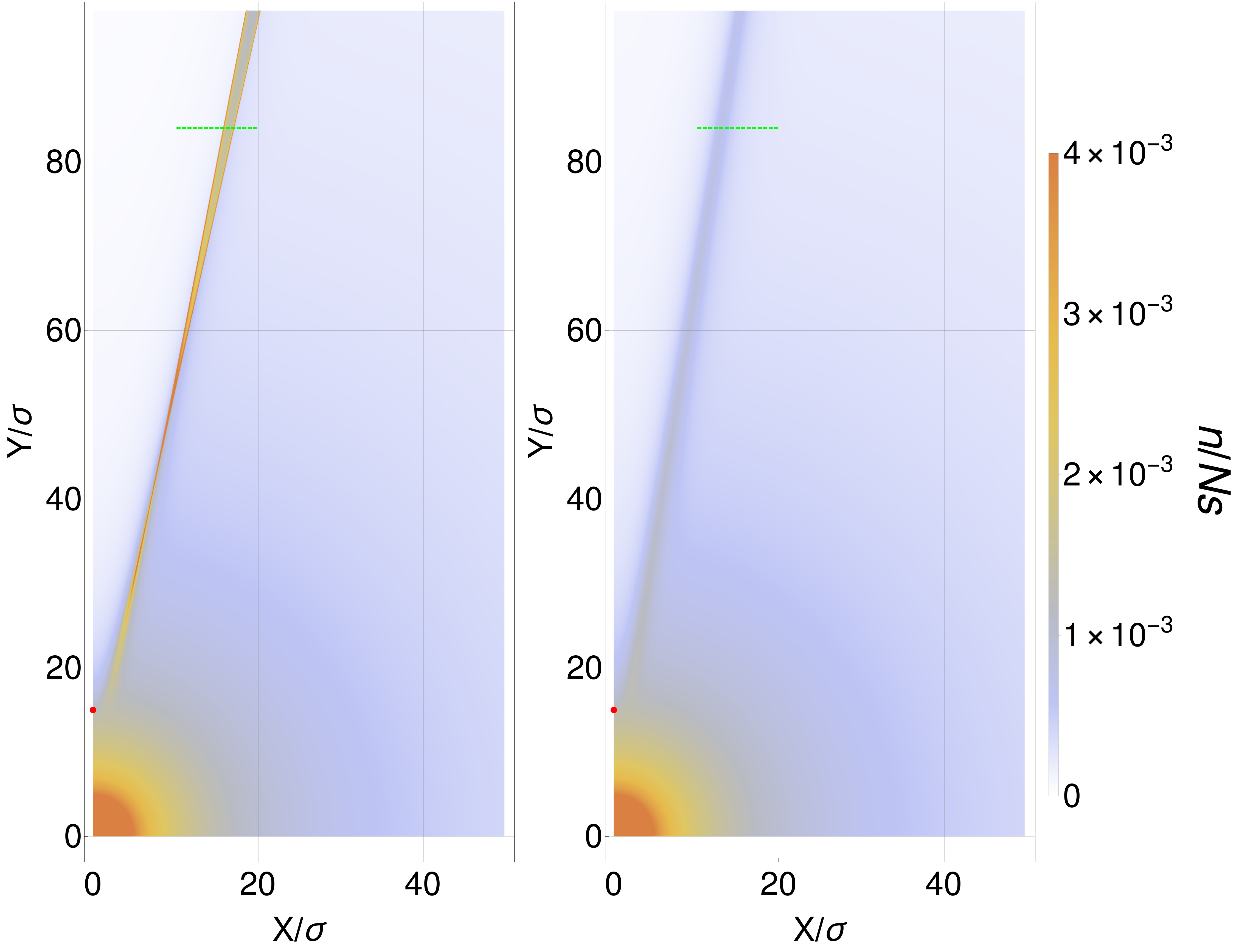}
\caption{Density of classical electron trajectories as a result of scattering from the same localized, repulsive Gaussian bump as in Fig.\ \ref{fig:zB}. The choice of parameters is the same as in Fig.\ \ref{fig:zB}, except the electron energy, that now is 5 (left) and 10 (right) $V_{0}$. The green dashed lines indicate the horizontal cuts considered in Fig.\ \ref{fig:hCP}.}
\label{fig:sBS}
\end{figure}
Figure \ref{fig:zB} displays the numerically calculated density of classical electron trajectories in the case of electrons scattered off of a radially symmetric Gaussian bump potential,
\begin{equation}
V \left ( r \right ) = V_{0} \exp \left ( -r ^{2} / 2 \sigma^{2} \right ),
\label{eq:potDefV}
\end{equation}
with $V_{0}$ the amplitude of the Gaussian bump, $r$ the distance from the center of the bump, and $\sigma$ the characteristic width of the bump (details of the numerics can be found in Appendix \ref{app:disGen}). Due to the radial symmetry of the potential, the figure is symmetric around the axis $x = 0$ and equivalent structures appear on either side of the bump. For simplicity, we study the case of a uniform angular distribution of electron trajectories starting at the QPC, and we have found that our results do not significantly depend on this choice. On either side of the figure, the formation of a well-collimated branch is clearly visible, while the rest of the background density of trajectories becomes fainter. In this figure, the energy of the electrons is tuned to be 8.23$V_{0}$, and the distance to the bump is taken to be 15$\sigma$, a parameter regime which we will later identify as a characteristic one for branch formation in this model.
A pattern similar to that of Fig.\ \ref{fig:zB} is obtained from a plot of the modulus squared of the wavefunction representing a plane-wave impinging on a localized bump, calculated by a multiple scattering expansion \cite{quantumCausticThesis}.

Figure \ref{fig:sBS} addresses the question of the branch stability with respect to a change of the electron's energy. The two panels display the density of trajectories for an energy smaller (left) and larger (right) than that of Fig.\ \ref{fig:zB}, while keeping all other parameters unchanged (only for $x > 0$, since the figures have axial symmetry around $x = 0$).  We see that the location of the branch is only slightly modified as we change the energy by a factor of two. Hence our toy model, in the case of a repulsive feature, exhibits stability against a relatively large shift in energy similarly to the experimental observation of Ref.\ \cite{braem2018stable}. We have found this stability to be independent of the precise shape of the potential, and also that it does not require radial symmetry. 
\begin{figure}
\centering
\includegraphics[width=\columnwidth]{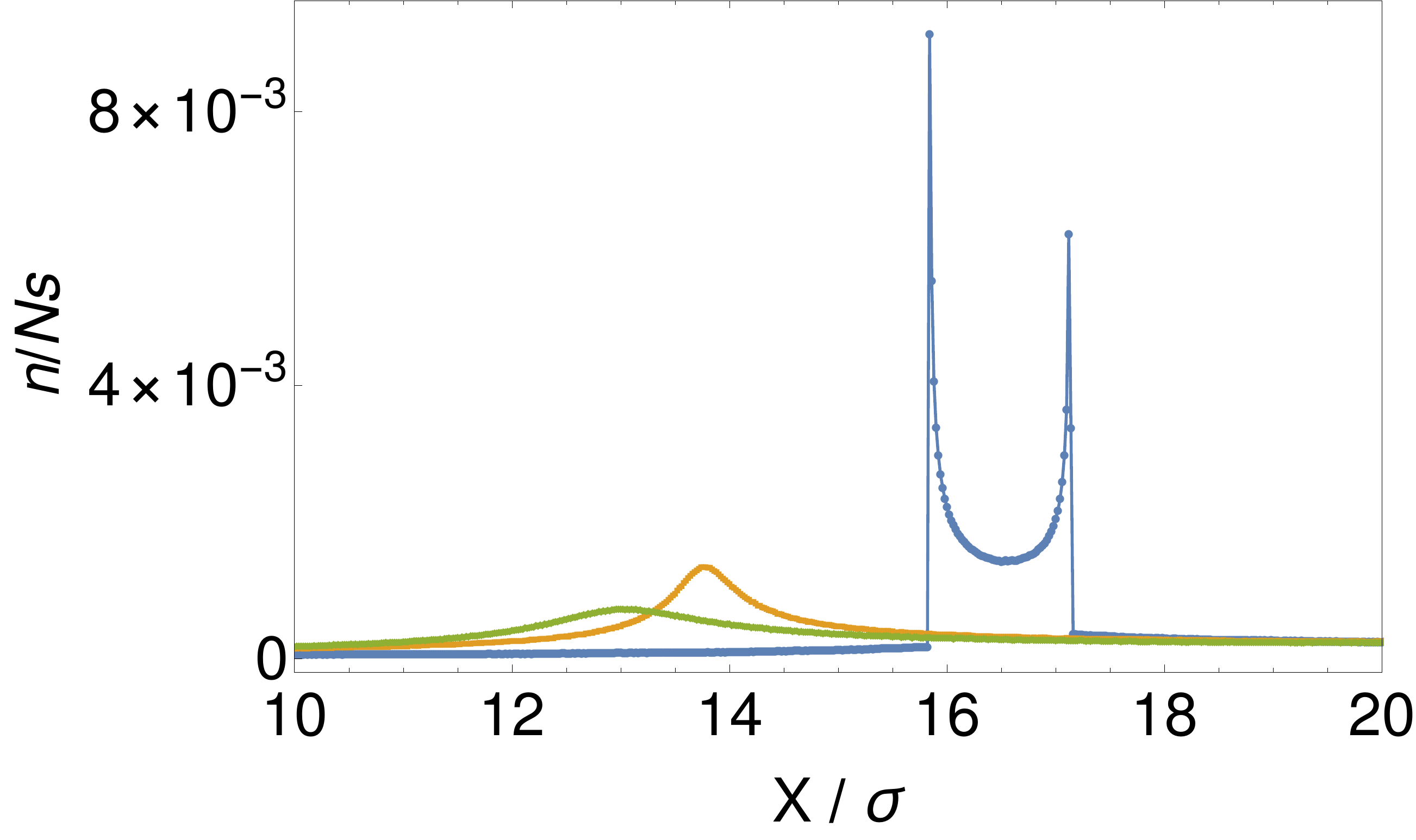} 
\caption{The density of trajectories across the horizontal cut displayed in Figs.\ \ref{fig:zB} and \ref{fig:sBS}, which is located at a vertical position of $y = 84 \sigma$. The blue, orange, and green data points correspond to energies of the scattered electrons equal to 5, 8.23, and 10 $V_{0}$, respectively.}
\label{fig:hCP}
\end{figure}
To obtain an intuitive understanding of branch formation and its energy stability, we display in Fig.\ \ref{fig:hCP} the spatial density of trajectories across a horizontal cut downstream with respect to the scatterer (indicated by the green dashed lines in Figs.\ \ref{fig:zB} and \ref{fig:sBS}) for different electron energies. For the case of the lowest energy, there are two closely-spaced divergent points marking the boundaries of the branch, and these divergences disappear for larger energies as we explain in the next sections. 
The trajectory densities of Fig.\ \ref{fig:hCP} show that the lateral shift of the branch is small on the scale of Figs.\ \ref{fig:zB} and \ref{fig:sBS}, which translates into the robustness of the SGM scans under energy changes. However, the values attained by the trajectory density present an important dependence on the electron energy. This observation is consistent with the variations of trajectory density seen in Figs.\ \ref{fig:zB} and \ref{fig:sBS}, and also with the SGM scans of Ref.\ \cite{braem2018stable}.

Notice that in Figs.\ \ref{fig:zB} and \ref{fig:sBS} there is a significant residual background flux of trajectories immediately outside of and far away from the branch. 
As we discuss in the next sections, once we go beyond the toy model, upon encountering another scattering center, the residual flux could induce the formation of another branch. In an experiment or a graphical representation with an appropriately adjusted resolution or threshold, these branches would be visible, while the residual background may not be.

\begin{figure}
\centering
\includegraphics[width=0.95\columnwidth]{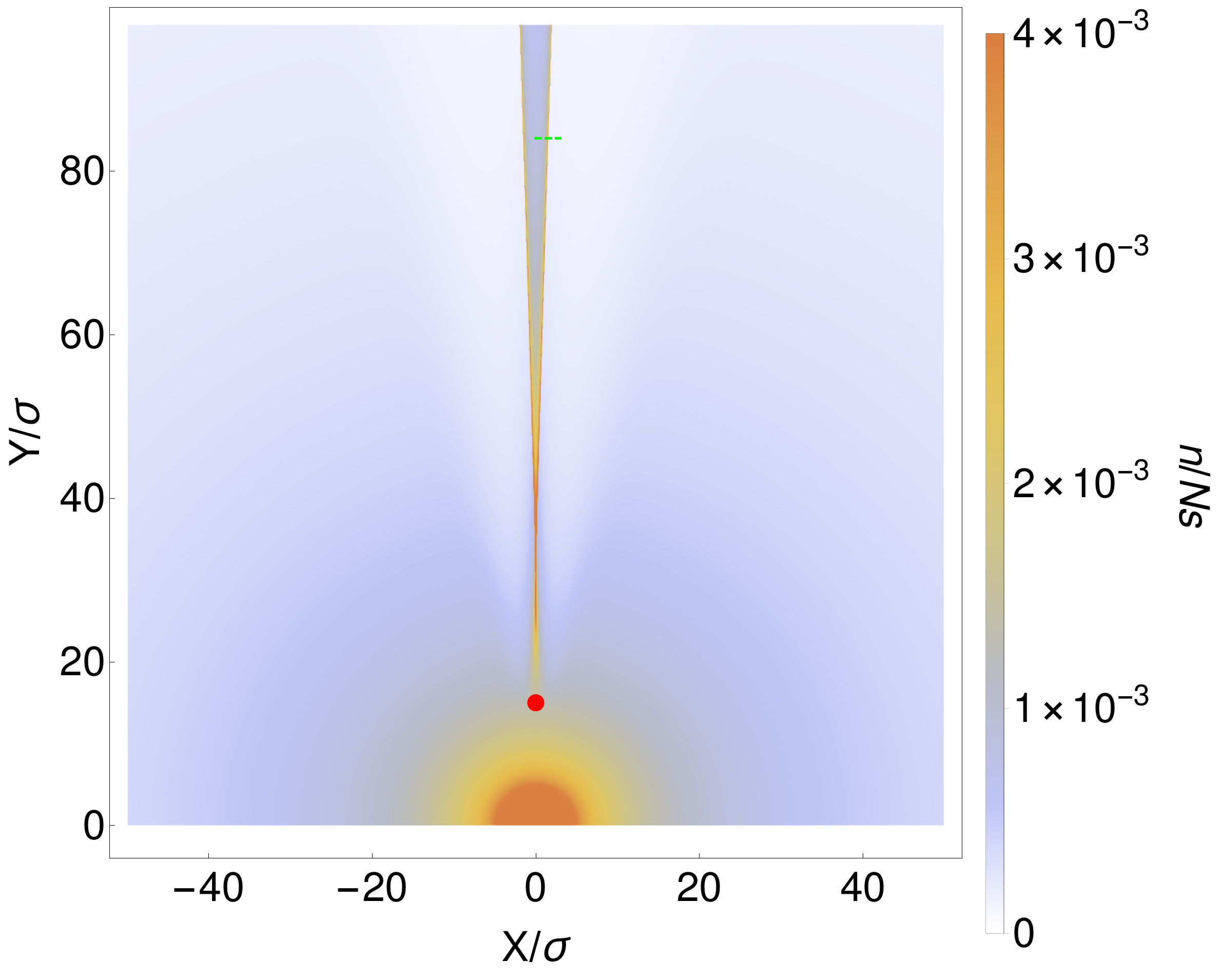} 
\caption{Density of classical electron trajectories as a result of scattering from a localized, attractive Gaussian bump (indicated by the red dot) with width $\sigma$ and height $V_{0}$. The distance between the center of the bump and the QPC is taken to be $a = 15 \sigma$, with the QPC taken to be at the origin. The choice of plotting parameters is the same as in Fig.\ \ref{fig:zB}, including the value of the electron energy $E = 8.23V_{0}$.}
\label{fig:zBN}
\end{figure}
\begin{figure}
\centering
\includegraphics[width=\columnwidth]{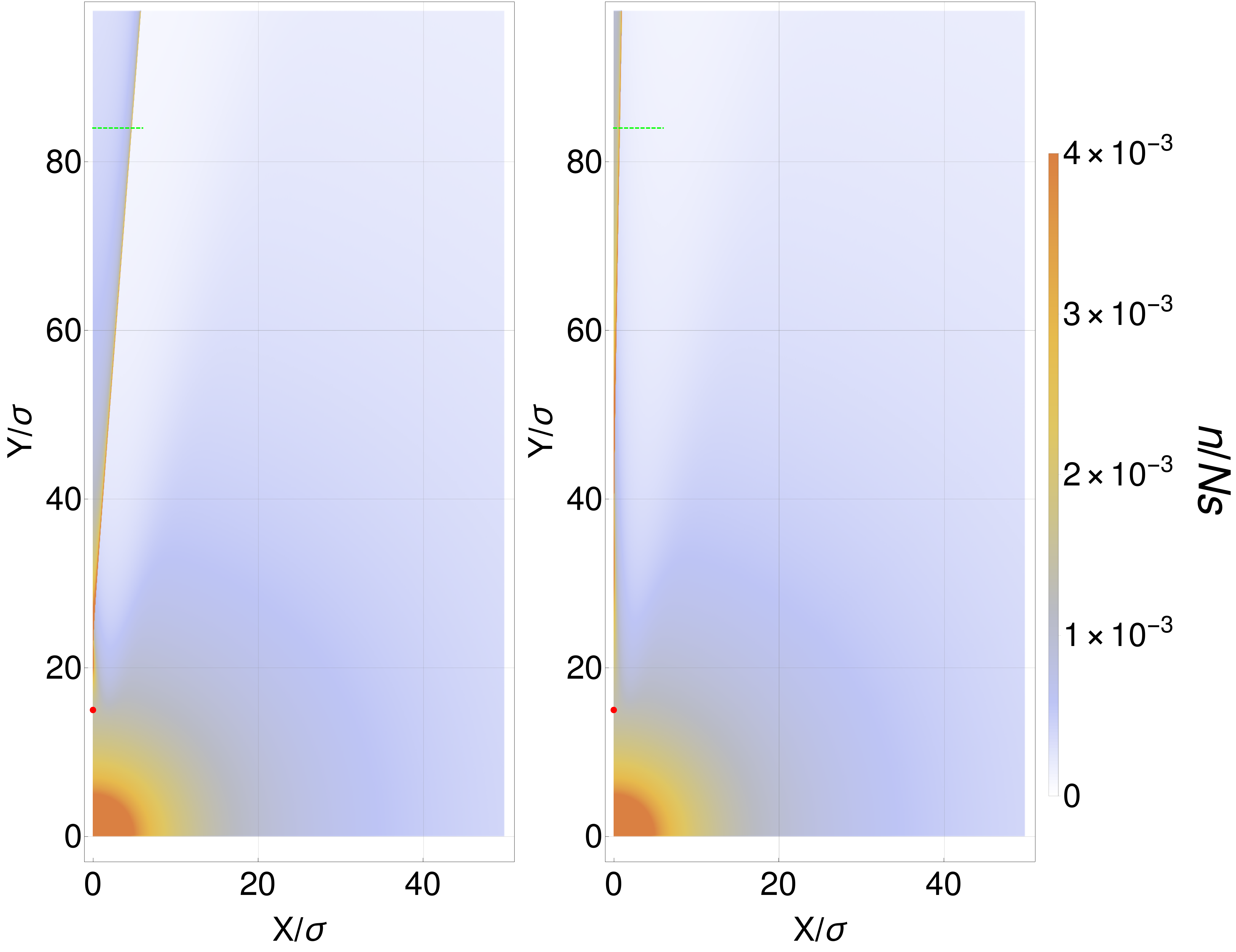}
\caption{Density of classical electron trajectories as a result of scattering from the same localized, attractive Gaussian bump as in Fig.\ \ref{fig:zBN}. The choice of parameters is the same as in Fig.\ \ref{fig:sBS}.}
\label{fig:sBSN}
\end{figure}

\subsection{The case of a local minimum}

In contrast with the previously discussed case of a repulsive feature, Fig.\ \ref{fig:zBN} demonstrates the density of classical electron trajectories after scattering from a weakly attractive localized feature, given by a radially symmetric, Gaussian profile. In this case, only one branch is observed, centered around the axis $x=0$, with two symmetric divergences marking the boundary of the branch. The classical trajectories are first focused by the attractive potential, before propagating outwards in these two symmetric divergences. 
The distance to the QPC is again taken to be 15 times the width of the localized feature, with the energy tuned to 8.23 times the amplitude of the localized feature. The two panels of Fig.\ \ref{fig:sBSN} display the density of trajectories for an energy smaller (left) and larger (right) than that of Fig.\ \ref{fig:zBN}, while keeping all the other parameters unchanged. Comparing the data of Figs.\ \ref{fig:zBN} and \ref{fig:sBSN} we observe that, in contrast with the case of a repulsive feature, there is less stability in the width of the branch for this energy regime when it is formed by an attractive feature, while the position of the branch is trivially stable at $x=0$. Additionally, the residual flux of trajectories is largely concentrated between the two divergences, as opposed to outside the region delineated by the divergences. 

\begin{figure}
\centering
\includegraphics[width=\columnwidth]{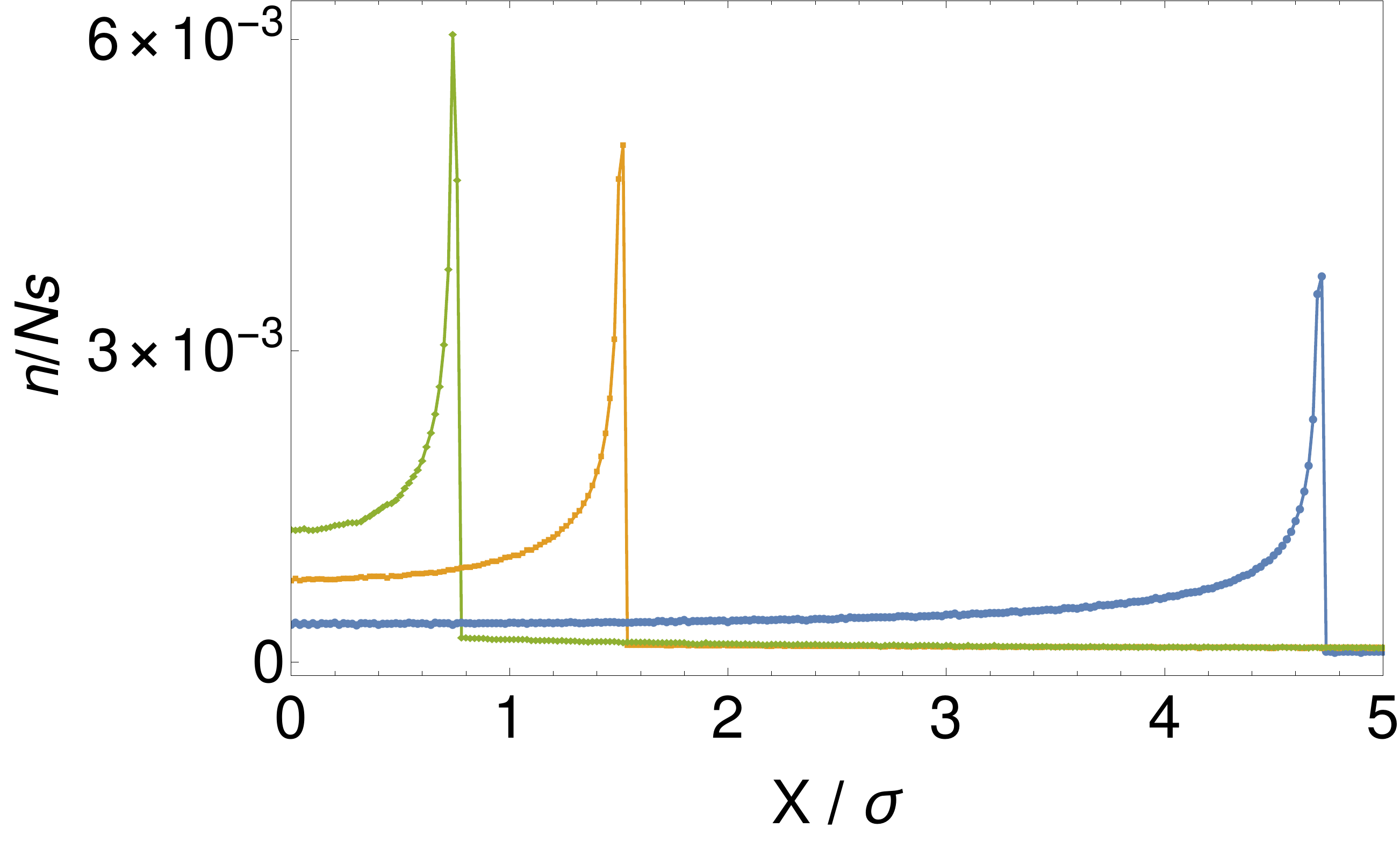} 
\caption{Density of trajectories across the horizontal cuts displayed in Figs.\ \ref{fig:zBN} and \ref{fig:sBSN}, which is located at a vertical position of $y = 84 \sigma$. The blue, orange, and green data points correspond to energies of the scattered electrons equal to 5, 8.23, and 10 $V_{0}$, respectively.}
\label{fig:hCN}
\end{figure}
Figure \ref{fig:hCN} displays the density of electron trajectories along the horizontal cuts of Figs.\ \ref{fig:zBN} and \ref{fig:sBSN}. Only the data for $x>0$ is shown, since the branches are symmetric.

\subsection{The role of energy in the branch formation for the case of a local maximum}
\label{sec:crit}

\begin{figure}
\centering
\includegraphics[width=0.7\columnwidth]{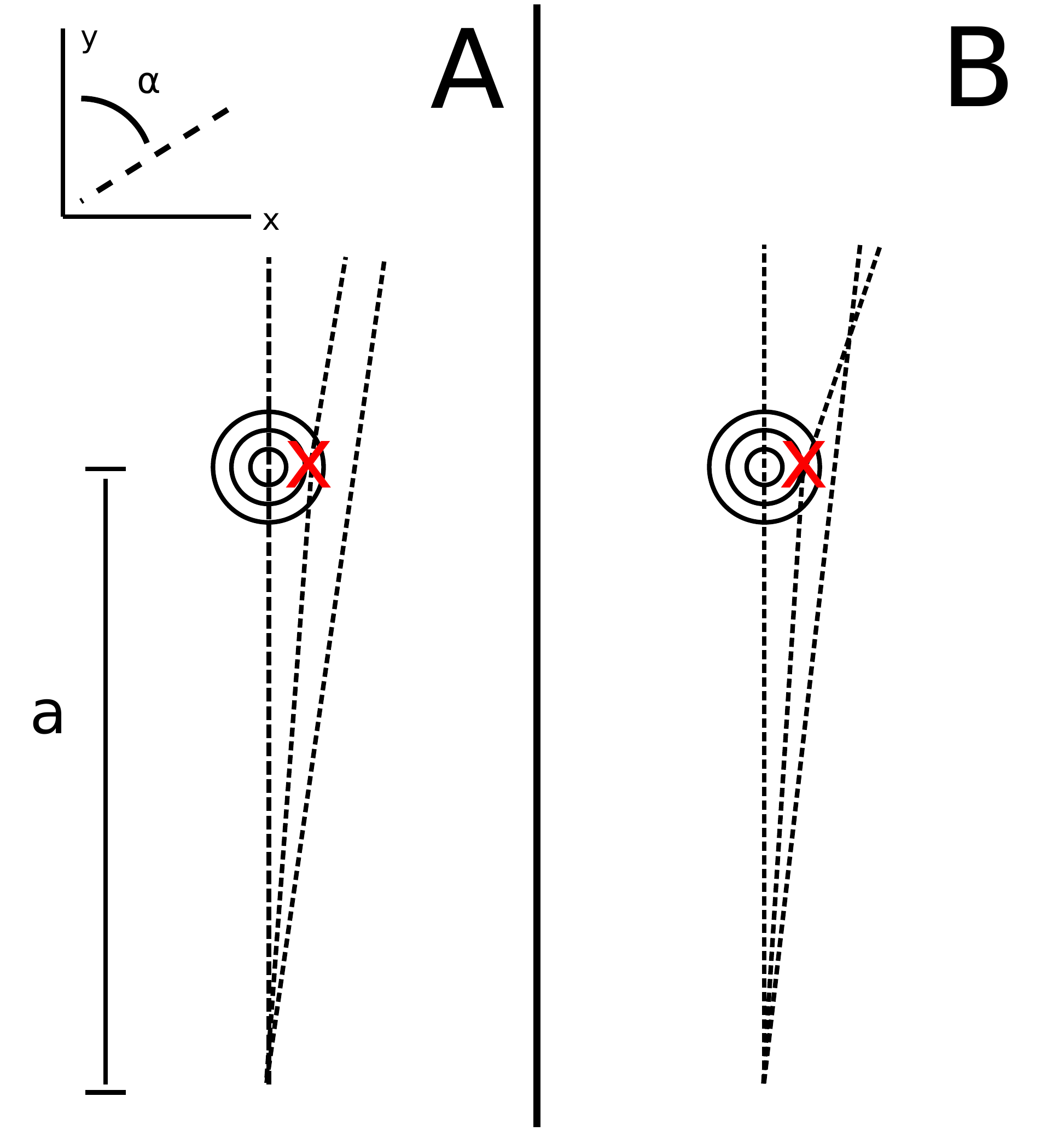}
\caption{A sketch of the scattering mechanism displaying classical trajectories at two energies $E_\mathrm{A} > E_\mathrm{B}$, emitted from a QPC and incident upon a localized, repulsive potential ``bump.'' Trajectories in the forward direction, as well as trajectories scattering at large angles, suffer minimal deflection. Trajectories at intermediate angles suffer some non-trivial scattering, depicted here in caricature by a red ``X.'' If the scattering function is not monotonic with initial angle, which will occur for sufficiently low energies, some trajectories will cross with each other behind the bump. The deflection of the scattered trajectories is exaggerated here for the sake of visual aide.}
\label{fig:qual}
\end{figure}
In order to understand the role that energy plays in the formation of branches, and to move towards a quantitative and qualitative explanation of the stability with respect to energy, we now consider in more concrete terms the scattering of classical trajectories as a function of energy in our toy model.

\begin{figure}
\centering
\includegraphics[width=0.9\columnwidth]{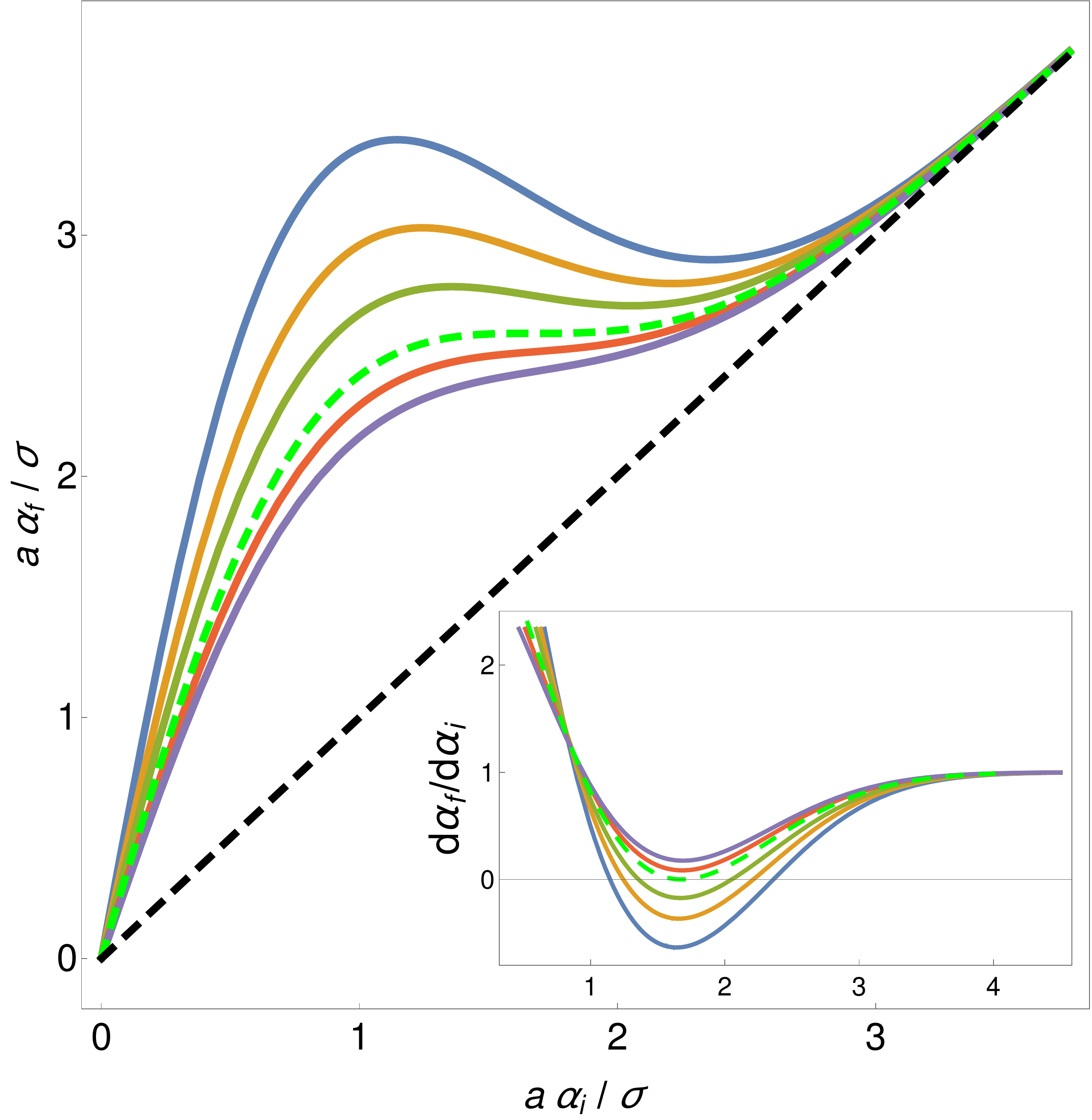}
\caption{Scattering function for electrons impinging upon a localized, repulsive bump as the energy is increased from top to bottom. As the critical condition is reached with lowering energy, the scattering function develops a zero derivative at exactly one point, and then develops a behavior which is not monotonic, with a zero derivative at two points. The green dashed curve represents the case in which the system is tuned precisely to the threshold energy. The inset to the figure displays the derivative of the scattering function, with the curves displayed in the opposite order, from top to bottom, as the main figure. These particular curves were computed for the example of a Gaussian bump at a distance $a=15\sigma$ from the QPC, though we find that this overall qualitative behavior does not depend on the detailed nature of the scattering potential. The corresponding energies, from top to bottom, are 5, 6, 7, 8.23, 9, and 10 times the height of the bump, with 8.23 being the threshold energy.}
\label{fig:sF}
\end{figure}
Figure \ref{fig:qual} displays a qualitative sketch of scattering for two different energies $E_\mathrm{A}$ and $E_\mathrm{B}$, with $E_\mathrm{A} > E_\mathrm{B}$. We measure angles $\alpha$, before and after the scattering event, with respect to the y-axis. In order to investigate the conditions for the sharp focusing of the classical trajectories by such a bump and how these conditions are affected by the electron's energy, we focus on the probability distribution $P_\mathrm{f}\left ( \alpha_\mathrm{f} \right )$ of final outgoing angles $\alpha_\mathrm{f}$, after scattering by the localized feature, that can be formulated as
\begin{equation}
P_\mathrm{f}\left ( \alpha_\mathrm{f} \right ) ~ = ~\int d \alpha_\mathrm{i}'~ P_\mathrm{i} \left ( \alpha_\mathrm{i}' \right )~\delta \left ( \alpha_\mathrm{f} - \mathcal{\aleph} \left ( \alpha_\mathrm{i}' \right ) \right ).
\label{eq:divProb}
\end{equation}
$P_\mathrm{i}$ is the distribution of initial angles $\alpha_\mathrm{i}$ and the scattering function $\mathcal{\aleph} \left ( \alpha_\mathrm{i} \right )$ computes the outgoing angle for a given initial angle. We restrict ourselves to the case of a radially symmetric bump, and thus limit our discussion to $ \alpha_\mathrm{i} > 0$. Using a standard identity for the Dirac delta function, we write
\begin{equation}
P_\mathrm{f}\left ( \alpha_\mathrm{f} \right ) =  \sum_{j} \frac{P_\mathrm{i}\left ( \widetilde{\alpha}_\mathrm{i}^{(j)} \right )}{ \left | \frac{d \mathcal{\aleph}}{d \alpha_\mathrm{i}} \left (  \widetilde{\alpha}_\mathrm{i}^{(j)} \right ) \right |  },
\label{eq:divProbUnif}
\end{equation}
with the sum taken over all initial angles $\widetilde{\alpha}_\mathrm{i}^{(j)} $ which satisfy $\mathcal{\aleph} \left ( \widetilde{\alpha}_\mathrm{i}^{(j)}  \right ) = \alpha_\mathrm{f}$. Thus, a divergence in the distribution of final angles, which can be associated with the formation of a branch, occurs when the scattering angle, as a function of the initial incident angle, presents a zero derivative. Large enhancements in the distribution of final angles, which can also be associated with the formation of a branch, occur when this derivative is small but non-zero. We will shortly see that these two categories of branches, corresponding to singular and non-singular behavior, lead to branching phenomena which would ultimately appear similar in an SGM experiment. Furthermore, a strict divergence of classical trajectories does not occur in the case of a finite-width QPC, and no divergence of the partial local density of states is expected in a fully quantum treatment of branching \cite{kaplan2002statistics,maryenko2012branching,vaniheller2003,heller2008refraction}, rendering the distinction even less crucial.

With such a criterion in mind, we now examine the qualitative form of the scattering function for our toy model shown in Fig.\ \ref{fig:sF}. The figure displays the shape of $\mathcal{\aleph} \left ( \alpha_\mathrm{i} \right )$ for several choices of $E / V_{0}$, the ratio of the electron energy to the amplitude of the potential bump. This figure is displayed for the case of a Gaussian bump, but its qualitative features are robust against the precise shape of the model potential. From top to bottom, the energies of the curves increase, with the middle dashed curve representing the energy at which the divergence condition is first satisfied. The scattering function depends on the initial angle through the impact parameter
\begin{equation}
b = a \sin \left ( \alpha_\mathrm{i} \right ) \approx a \alpha_\mathrm{i}~,
\label{eq:imparamdef}
\end{equation}
where the approximation is valid in the limit of small angle scattering. Assuming our localized feature is defined in terms of a characteristic width $\sigma$, the final scattering angle becomes a function of the combination $a \alpha_\mathrm{i} / \sigma$, as it is displayed in Fig.\ \ref{fig:sF}.

Since the height of the potential bump is assumed to be lower than the electron energy, trajectories which approach the center of the bump head-on at zero initial angle suffer minimal deflection, and scatter largely in the forward direction, corresponding to zero final angle. In the simplifying case of a symmetric bump, there will be zero deflection, with $\alpha_\mathrm{f} = \alpha_\mathrm{i} = 0$. At small angles away from head-on scattering, the trajectory will begin to suffer some outward angular deflection, so that
\begin{equation}
\frac{d \alpha_\mathrm{f}}{d \alpha_\mathrm{i}} > 1~;~\alpha_\mathrm{f} > \alpha_\mathrm{i}.
\end{equation}
At larger angles, where the potential vanishes, the deflection will again be essentially zero, so that the scattering function will asymptotically approach $\alpha_\mathrm{f}  = \alpha_\mathrm{i}$. In this case, the derivative of the scattering function is unity. Whether or not there will be a zero derivative at some point between these two limiting cases depends on whether the scattering function remains monotonically increasing at intermediate angles. For a given scattering bump, whether this condition holds will depend on the energy of the incident electron. Figure \ref{fig:qual} displays the possible cases, which we will now address in turn.

Case A in Fig.\ \ref{fig:qual} represents the situation in which the electron energy is high enough that the scattering function always remains monotonic. Trajectories with larger initial angles scatter to larger final angles, and trajectories never cross. This corresponds to the lower curves in Fig.\ \ref{fig:sF}. In case B, the energy is low enough that the intermediate angles suffer very large deflection. Trajectories at smaller initial angle cross with trajectories at larger initial angle behind the bump. In this case, the final scattering angle $\alpha_\mathrm{f}$ as a function of incident angle $\alpha_\mathrm{i}$ must reach a local maximum, then decrease to a local minimum, before asymptotically approaching $\alpha_\mathrm{f}  = \alpha_\mathrm{i}$. Such a non-monotonic behavior of the scattering function leads to two points with zero derivative. This corresponds to the upper curves in Fig.\ \ref{fig:sF}. The transition between these two scenarios occurs at a critical threshold energy at which the derivative of the scattering function is zero at precisely one initial angle. This corresponds to the central, dashed scattering curve in Fig.\ \ref{fig:sF}, and the choice of parameters in Fig.\ \ref{fig:zB}.

At each initial angle where the scattering function exhibits a zero derivative, the outgoing probability distribution diverges on the corresponding outgoing angle. Such a divergence that occurs due to the vanishing of $ d \mathcal{\aleph} / d \alpha_\mathrm{i}$ is the requirement necessary for the sharp accumulation of trajectories, which in case B occurs at two points, and at the crossover between cases A and B occurs at precisely one point. However, we will shortly see that in case B, these two points can be identified as forming only a single branch.

The condition for divergence formation we have derived here for a point-like QPC, that the derivative of the final scattering angle must be zero with respect to the initial scattering angle, can be shown, under the assumption of weak scattering, to be equivalent to the more traditional condition \cite{stockmann2006quantum} for the formation of a caustic,
\begin{equation}
\left | \frac{\delta q\left ( t \right )}{\delta p_\mathrm{i}} \right | = 0,
\label{eq:caustic}
\end{equation}
where $\delta q \left ( t \right )$ is the position of some scattered electron at some fixed time $t$, and $p_\mathrm{i}$ is the initial momentum of the electron trajectory. We elaborate on this equivalency in Appendix \ref{sec:caustEq}.

\subsection{The role of energy in the branch formation for the case of a local minimum}

\begin{figure}[t!p]
\centering
\includegraphics[width=0.7\columnwidth]{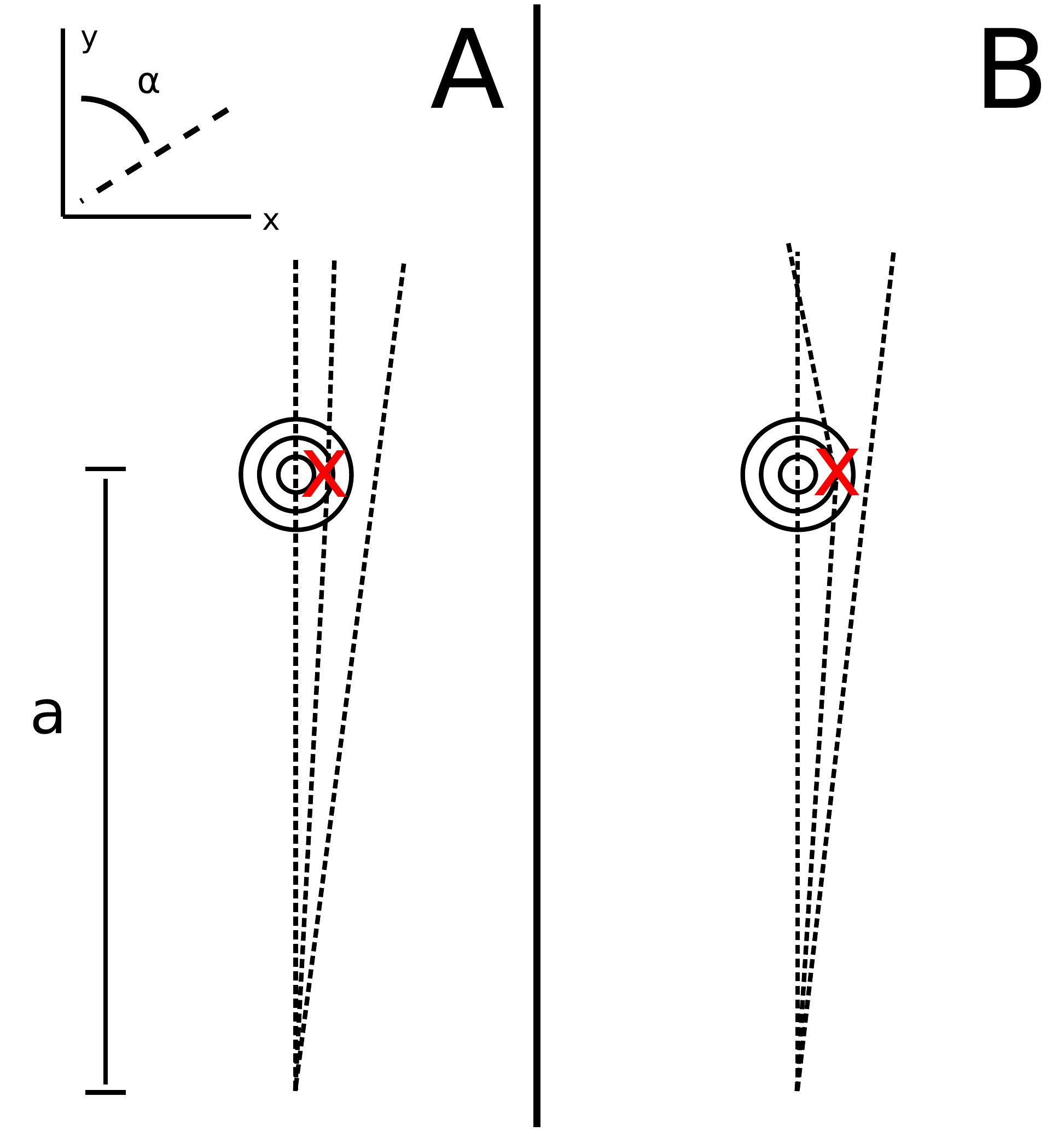}
\caption{The same qualitative sketch as Fig.\ \ref{fig:qual}, now for the case of an attractive potential ``dip.''}
\label{fig:qualN}
\end{figure}

Turning to the case of an attractive feature, Figs.\ \ref{fig:qualN} and \ref{fig:sFN} display the scattering mechanism and scattering function, respectively. Again, there are two qualitatively distinct cases, corresponding to low and high energies. In both cases, $\alpha_\mathrm{f} = \alpha_\mathrm{i}$ at zero initial scattering angle, as well as large initial scattering angle. At intermediate angles, the scattering function is again non-trivial. At high energies, the scattering function is monotonic - while it is always necessarily bounded above by $\alpha_\mathrm{f} = \alpha_\mathrm{i}$, the scattering function is still strictly increasing. At small angles the derivative of the scattering function is less than unity, yet still positive, and approaches unity monotonically at large angle. This corresponds to case A in Fig.\ \ref{fig:qualN}.

At lower energies, corresponding to case B, the scattering function is no longer monotonic. Trajectories at small initial angle are deflected more strongly as the angle is increased, corresponding to a scattering function with negative derivative. At some non-zero angle, the scattering function reaches a minimum, corresponding to maximum deflection in the negative direction, and hence obtains a zero derivative. As can be seen in the scattering function, this occurs for only one angle, as opposed to two angles for the case of the repulsive feature. This angle shifts towards zero as the energy is raised, and the case of intermediate energy occurs when the derivative of the scattering function is zero at precisely zero initial scattering angle.

In this case, the presence of a zero derivative at only one angle at low energies corresponds to the single branch centered around the axis $x=0$, whose width varies considerably with energy, yet whose position is trivially fixed.

\subsection{Stability of the branching patterns with energy}\label{sec:enStab}

\begin{figure}[t!p]
\centering
\includegraphics[width=0.9\columnwidth]{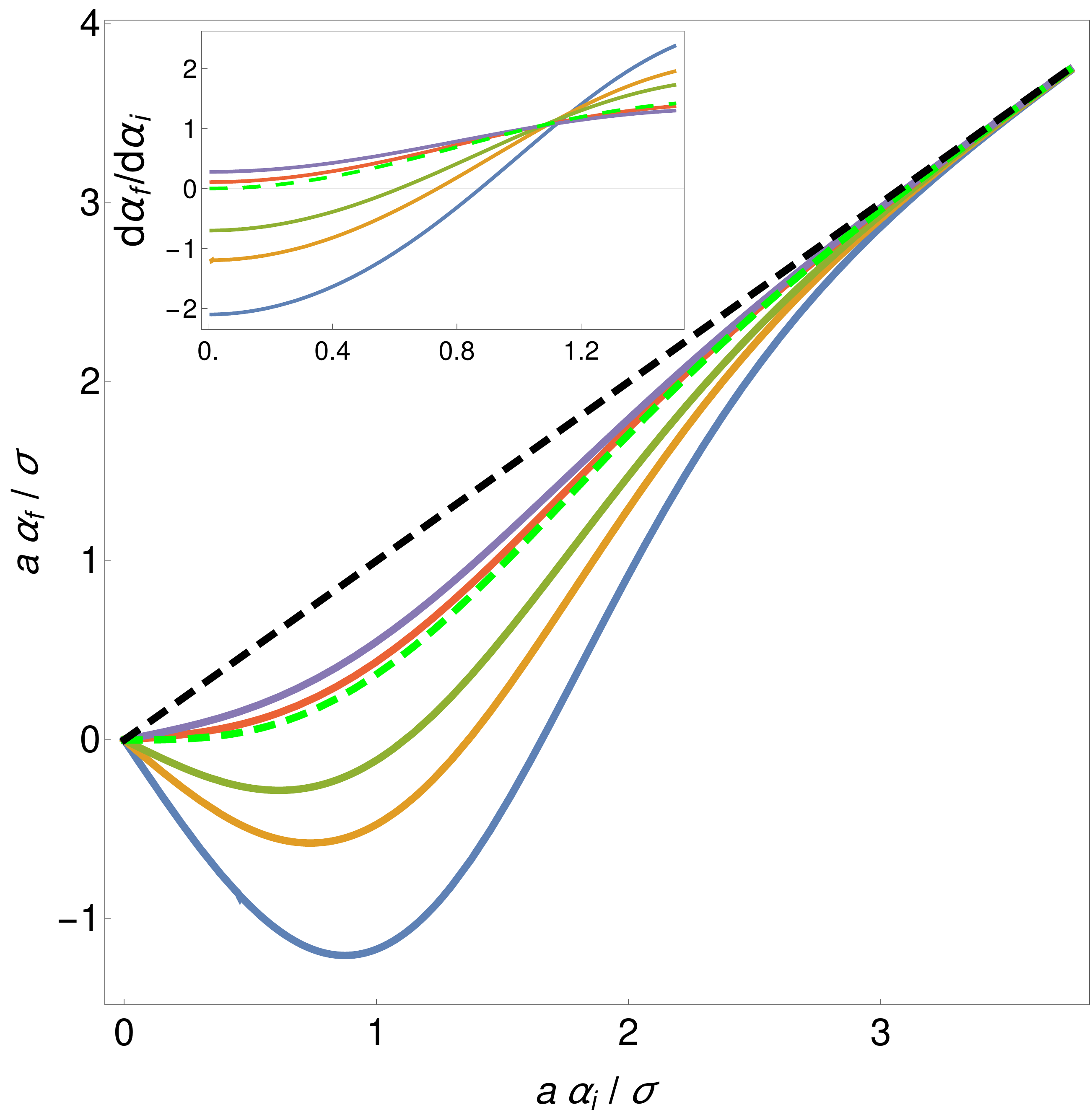}
\caption{Scattering function for electrons scattering from a localized, attractive bump as the energy is decreased from top to bottom. As the critical condition is reached with lowering energy, the scattering function develops a zero derivative at the origin, and then develops a behavior which is not monotonic, with a zero derivative that shifts to larger angle with lower energy. The green dashed curve represents the case in which the system is tuned precisely to the threshold energy. The inset shows the derivative of the scattering function, with the curves displayed in the same order, from top to bottom, as the main figure. These particular curves were computed for the example of a Gaussian bump at a distance $a=15\sigma$ from the QPC, though we find that this overall qualitative behavior does not depend on the detailed nature of the scattering potential. The corresponding energies, from top to bottom, are 25, 20, 17.75, 10, 7.5, and 5 times the height of the bump, with 17.75 being the threshold energy.}
\label{fig:sFN}
\end{figure} 
While the qualitative nature of the scattering functions in Figs.\ \ref{fig:sF} and \ref{fig:sFN} appears to vary considerably over the range of displayed energies, the structure of the branches in Figs.\ \ref{fig:sBS} and \ref{fig:sBSN} looks remarkably stable. We now investigate the underlying mechanism for this energy stability in our toy model. This stability crucially depends on the qualitative shape of the scattering function, and how it evolves with changing energy. 

We begin with the case of a repulsive feature. At higher energies, while there is no point at which the scattering function is zero, we can see in Fig.\ \ref{fig:sF} that there is a range of angles over which the derivative of the scattering function comes very close to zero, resulting in a peak of $P_\mathrm{f}$ in Eq.\ \eqref{eq:divProbUnif}, though not a strict divergence. In fact, the inset to Fig.\ \ref{fig:sF} demonstrates that the point at which the derivative reaches a minimum, and hence comes closest to zero, is effectively the same as the initial angle at which the derivative first crosses zero, $\alpha^{*}_\mathrm{i}$, as the energy is lowered. 

As the energy of the trajectories is lowered below the threshold energy, there are two points at which there is a zero derivative, which move further apart from each other with decreasing energy. However, we find that the point between these two locations remains relatively fixed at $\alpha^{*}_\mathrm{i}$, so that the central impact parameter between these two angles, and hence the central location of branch formation, remains stable.

Below the critical energy, the shape of the scattering function, with a peak at smaller angles, and a valley at larger angles, causes the scattered trajectories in the two divergences to bend inwards slightly towards a central point, centered around trajectories which possess an initial scattering angle of $\alpha^{*}_\mathrm{i}$. An experimental probe like SGM, having a finite spatial resolution, is not able to distinguish the case of branch formation which is precisely at the threshold energy, from the case of branch formation due to two slightly inward bending divergences that converge towards the same central angle. Hence, a branch, originating from the same central impact parameter, and hence the same initial point in space, will still be visible.

For this reason, below the threshold energy, there is in fact a somewhat stronger divergence of trajectories in physical space, in the near-field region, as the trajectories corresponding to the two divergent points will cross at some finite distance from the scattering center. However, in a more realistic model of disorder, this distinction is not crucial, as subsequent scattering off of the disorder potential will render any questions of propagation over very long distances irrelevant. 

In fact, in Fig.\ \ref{fig:sF}, we observe that for a range of energies around the threshold energy, the scattering function becomes very flat over a broad range of initial scattering angles, centered around the critical angle where there is a zero derivative. Thus, a large number of trajectories scatter into nearly the same outgoing angle, and thus these trajectories propagate in parallel. However, they do so with different initial scattering angles, and hence a non-zero range of impact parameters. The outgoing branch thus has a finite width, which translates into the visibility and robustness of branches as seen by measurements with finite resolution.  

However, as the energy is varied, an examination of Fig.\ \ref{fig:sF} demonstrates that while the initial angle $\alpha^{*}_\mathrm{i}$ at the center of the branch remains relatively stable, the final outgoing angle corresponding to this initial angle changes moderately (with stronger deflection for lower energies). Thus, while the branch may initially form in the same location in space, the angular orientation of the outgoing branch may change slightly with energy, resulting in a slightly displaced branch. This phenomenon is indeed observed in our toy model of branch formation in Fig.\ \ref{fig:sBS}.

To elucidate the magnitude of the previously discussed phenomenon, we consider the case of a Gaussian bump with amplitude 1~meV and width 50~nm, placed 750~nm from the QPC, chosen to resemble realistic experimental values. For energies 6, 7, and 8~meV, just below the threshold energy of 8.23~meV, the central angle halfway between the two points with zero derivative is given as 0.1152, 0.1137, and 0.1123 radian, with corresponding outgoing angles of 0.1935, 0.183, and 0.1746 radian. Thus, over a change of 2~meV, the outgoing angle of the branch deviates by 0.0189 radian, equal to an approximately ten percent shift in the outgoing angle, in terms of the initial scattering angle, and would lead to the branch becoming displaced by 18.9~nm over a propagation length of 1 micrometer. This sort of shift in physical space is consistent with those seen in experiments \cite{braem2018stable}.

In contrast, we see that when considering the case of an attractive feature, there is only one minimum of the scattering function, and its location moves substantially as we vary the energy of the scattered electrons. Again, this behavior has been found to be generic, and does not rely on the detailed choice of potential. This leads to one branch which has a highly variable width, but is trivially fixed in space in the forward direction. In the following section, we will understand the reason for this generic qualitative difference.

\section{Generic Nature of the Energy Stability for a Localized Feature}\label{sec:enStabGen}

For both, the repulsive and attractive localized features, we have found a generic behavior, which does not depend on the detailed shape of the potential, so long as it meets the basic requirements outlined at the beginning of Sec.\ \ref{sec:mech}. Additionally, while we have studied the simplified case of an isolated Gaussian bump, we note that \textit{any localized feature of the potential where the scattering function develops a zero derivative will lead to the formation of branches that are quite robust with respect to the electron energy.} A saddle point, or any other sloping feature of the potential presenting a similar scattering behavior will result in similar branching. We emphasize that the necessary accumulation of classical trajectories occurs when the energy is higher, possibly significantly, than the amplitude of the scattering potential. The classical trajectories are not being ``guided'' by any sort of valley in the potential. Rather, the trajectories are being weakly scattered by the features of the potential, in such a way that a large number of trajectories are being deflected with almost the same outgoing angle, leading to an increase in the density of trajectories in that direction.

As our explanation for the stability of the branching structure relies on the qualitative shape of the scattering function when the energy is varied, we now discuss the scattering in the generic case, not restricted to the particular case of our toy model with a Gaussian bump. As a result of energy and angular momentum conservation, it is a straight-forward exercise (see Appendix \ref{app:scatterApprox}) to find that the final scattering angle for an electron scattering off of a localized, radially symmetric potential $V$ is given according to
\begin{equation}
\alpha_\mathrm{f} = \alpha_\mathrm{i} + \pi - I \left ( b,E,V \right ),
\label{eq:scattAngMain}
\end{equation}
with
\begin{equation}
I \left ( b,E,V \right ) \equiv 2\int_{1}^{\infty} \frac{ds/s^{2}}{\sqrt{\lambda^{2}-1/s^{2}-\lambda^{2}~V\left (\lambda b  s \right )/E}},
\label{eq:intDefMain}
\end{equation}
where $E$ is the energy of the classical trajectory, $b$ is the impact parameter from Eq.\ \eqref{eq:imparamdef}, and $\lambda$ satisfies
\begin{equation}
1-1/\lambda^{2} - V \left ( \lambda b \right ) / E = 0.
\label{eq:paramDef1}
\end{equation}

While this is indeed the expression we have used to generate the plots of $\mathcal{\aleph} \left( \alpha_\mathrm{i} \right )$ in Fig.\ \ref{fig:sF}, it does not lend itself to a simple interpretation in the context of the energy stability. To gain more insight into this matter, we will simplify our problem by making the approximation that the electron energy is much higher than the typical potential energy, and suffers only weak deflection. We have checked that this approximation is extremely good for the parameter regimes we are interested in. 

We leave the details of the derivation to Appendix \ref{app:scatterApprox}, where we find the expression for the scattering function in this limiting case
\begin{equation}
\alpha_\mathrm{f}  \approx \alpha_\mathrm{i}  -\frac{1}{E}f_{a,V} \left ( \alpha_\mathrm{i} \right ).
\label{eq:impulseResult}
\end{equation}
We have defined
\begin{equation}
f_{a,V} \left ( \alpha_\mathrm{i} \right )~ \equiv~ \frac{a \alpha_\mathrm{i}}{\sigma}\int_{1}^{\infty}\frac{d\xi}{\sqrt{\xi^{2}-1}}\left. \frac{\partial V}{\partial \left ( r / \sigma \right )} \right |_{r ~=~ a \alpha_\mathrm{i}\xi},
\label{eq:fDef}
\end{equation}
which depends on the distance to the scattering center and the shape and width of the potential, but not the energy of the incident electrons. We have assumed that $V$ is defined in terms of a characteristic width, and only depends on the ratio $r / \sigma$. Our condition for a zero derivative of the scattering function then becomes
\begin{equation}
f'_{a,V} \left ( \alpha_\mathrm{i} \right ) - E = 0.
\end{equation}

In addition to being derivable through an impulse approximation, Eq.\ \eqref{eq:impulseResult} can also be obtained by simply performing a Taylor series expansion of the integral expression in Eq.\ \eqref{eq:intDefMain} in the parameter $V / E$. 

\begin{figure}
\centering
\includegraphics[width=0.9\columnwidth]{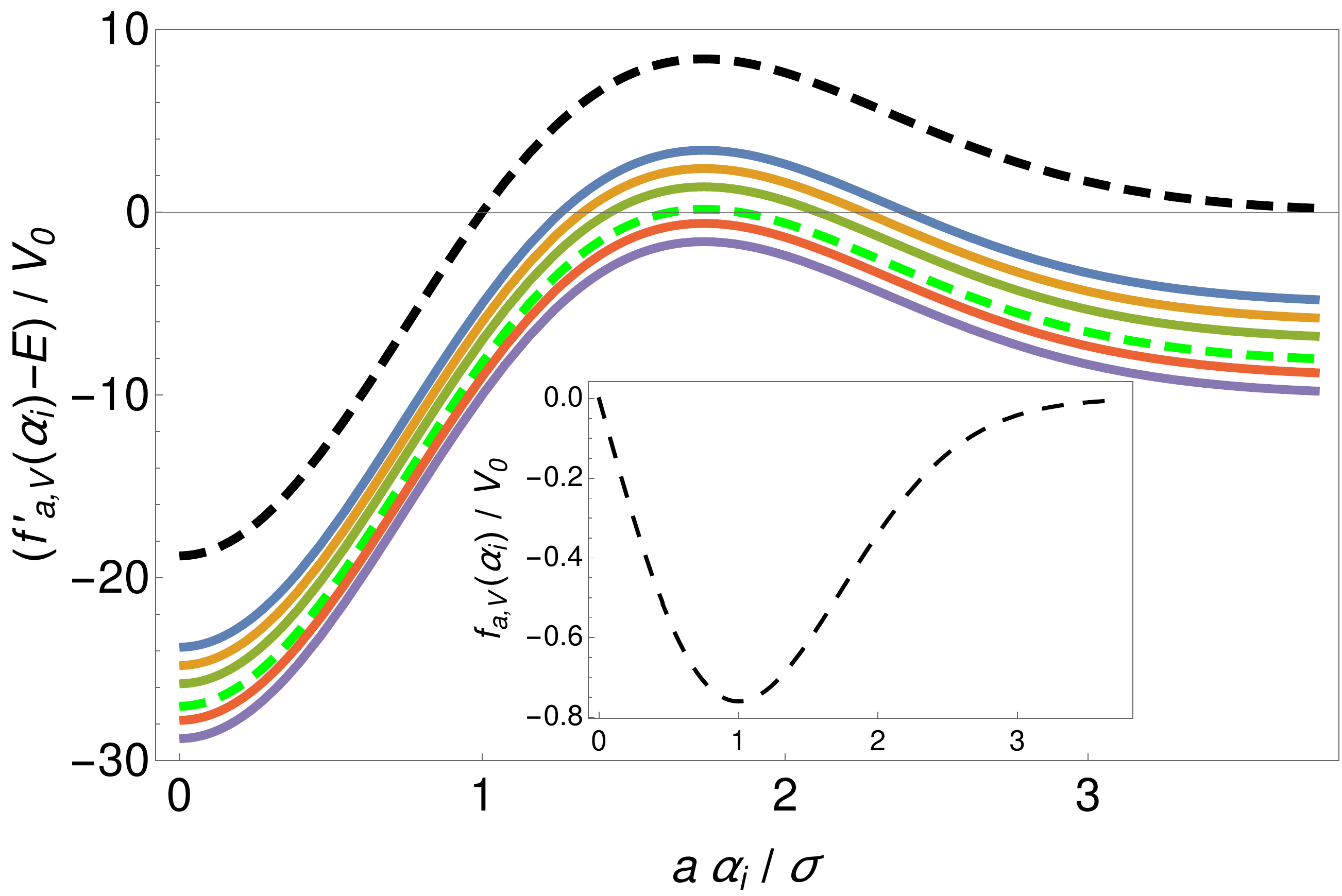}
\caption{The function $f'_{a,V}-E$ in the impulse approximation, generated for the case of a repulsive Gaussian bump with width $\sigma$, placed at a distance $a=15\sigma$ from the QPC. The energies $E$, from top to bottom, are 5, 6, 7, 8.23, 9, and 10 $V_{0}$. The inset to the figure displays the function $f_{a,V}$.}
\label{fig:sS}
\end{figure}
Beginning with the case of a repulsive feature, Fig.\ \ref{fig:sS} displays $f'_{a,V} - E$ for several choices of energy, while the inset displays the function $f_{a,V}$. The case of a Gaussian bump with width $\sigma$ and distance $a$ from the QPC has been chosen for illustration purposes. The upper-most curve in Fig.\ \ref{fig:sS} represents the original function $f'_{a,V}$. The key observation regarding the shape of $f'_{a,V}$ in Fig.\ \ref{fig:sS} is that it reaches a local maximum, with a value independent of the energy. This general feature must always be present for any reasonable choice of potential, as we can easily argue on the basis of Eq.\ \eqref{eq:fDef}. For small angles, the prefactor of $\alpha_\mathrm{i}$ in the definition of $f_{a,V}$ will guarantee
\begin{equation}
f_{a,V} \left ( 0 \right ) = 0.
\end{equation}
As $\alpha_\mathrm{i}$ is increased, and the potential decreases away from its maximum, we have
\begin{equation}
V' < 0 ~\Rightarrow ~f_{a,V} < 0.
\end{equation}
The value of $f_{a,V}$ will reach a minimum where $\alpha_\mathrm{i}$ is tuned to some intermediate value such that the integral over $V'$ is largest in magnitude, before approaching zero again at large angles. This is indeed the qualitative behavior seen in the inset to Fig.\ \ref{fig:sS}. The derivative of this function must reach some maximum positive value as the magnitude of the scattering approaches zero again, which is in fact observed in Fig.\ \ref{fig:sS}. 

As a result, adjusting the energy modifies the condition for a zero derivative, $f' - E = 0$, in a way that simply corresponds to translating the function $f'$ vertically. For large $E$, this function is shifted entirely below the axis, while for low enough $E$, it intersects the origin at two points. Since any analytic function will be symmetric around a local maximum for small enough deviations, this explains the symmetry of the scattering function geometry observed in Sec.\ \ref{sec:enStab}, which was crucial for explaining the stability of $\alpha^{*}_\mathrm{i}$ with respect to a change in the electron energy. In fact, one could take the Taylor expansion of the function $f'$ as fundamentally defining the energy range over which the branches formed by a repulsive feature should be stable, as the magnitude of the third order term in the expansion will measure the extent to which the function is symmetric around its local maximum.

\begin{figure}
\centering
\includegraphics[width=0.9\columnwidth]{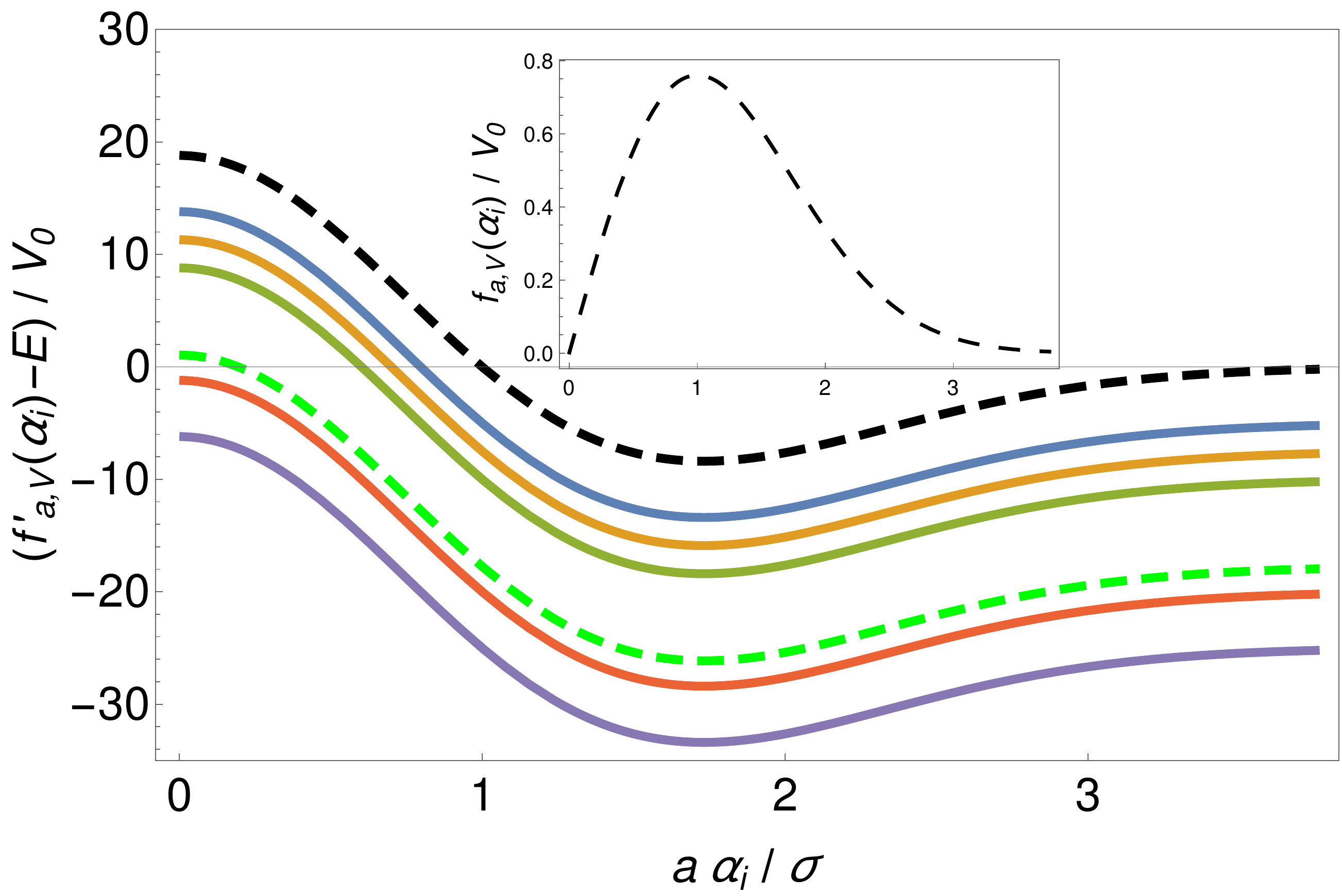}
\caption{The function $f'_{a,V}-E$ in the impulse approximation, generated for the case of an attractive Gaussian bump with width $\sigma$, placed at a distance $a=15\sigma$ from the QPC. The energies $E$, from top to bottom, are 5, 7.5, 10, 17.75, 20, and 25 $V_{0}$. The inset to the figure displays the function $f_{a,V}$.}
\label{fig:sSN}
\end{figure}
In contrast to this, Fig.\ \ref{fig:sSN} and its inset display the functions $f'_{a,V} - E$ and $f_{a,V}$, respectively, for the case of an attractive feature of the same width and shape. In this case, the functions $f_{a,V}$ and $f'_{a,V}$ merely acquire an overall minus sign. As a result, the qualitative behavior of $f'_{a,V}$ is such that it is positive at zero angle, obtains some minimum negative value at some non-zero scattering angle, and then eventually asymptotes to zero from below. In order for the function $f'_{a,V} - E$ to cross the horizontal axis at two points, we would require the total energy of the electrons to be negative, an unphysical result, as the electrons are not bound by the disorder potential. As we shift the function  $f'_{a,V} - E$ downwards with some positive energy $E > 0$, there is only ever one point which crosses the horizontal axis, the location of which is not fixed in place by any symmetry principle. As we eventually lower the energy far enough that the entire curve is below the horizontal axis, we are left with only the trivial case of an enhancement of trajectories at zero initial angle. Thus, we see that there is only one branch formed in the case of an attractive feature, and its width varies considerably with respect to energy. At sufficiently high energies, there is no divergence in the density of trajectories, only a non-singular enhancement. 

We thus see that the above-discussed geometric features of the scattering function are completely generic, so long as the general requirements on the potential outlined in Sec.\ \ref{sec:mech} are satisfied, and the scattering of electrons is sufficiently weak. 

While we have described the mechanism of branch formation due to a local feature in the context of electrons with varying energies and fixed distance to the QPC, there is also a complementary picture, in which the energy is kept fixed, and the distance to the QPC is varied. We can repeat our argument with cases A and B representing bumps which are at different distances, but with the same incident electron energy. We can estimate (see Appendix \ref{app:scatterApprox}) that for weak deflection, the strong accumulation criterion will be satisfied so long as
\begin{equation}
a |V_{0}| / \sigma E  \gtrsim  1,
\label{eq:impulse11}
\end{equation}
where $a$ is again the distance from the QPC to the bump, $\sigma$ is the characteristic width of the bump, $E$ is the energy of the electrons, and $V_{0}$ is the height of the potential. Thus, the width of the bump, in combination with the amplitude of the localized potential (with respect to the energy), sets the critical distance required for the formation of branching from a single potential feature to $a_\mathrm{c} \propto \sigma E/|V_0|$. Note that this condition holds for both repulsive and attractive bumps.
For the typical distance of the first caustic formation in classical flow through a weak random potential, a slightly different scaling, $a_\mathrm{t}/\xi \propto (E/v_0)^{2/3}$, where $\xi$ is the correlation length and $v_0$ the amplitude of the random potential, has been derived from diffusive motion in the transverse direction \cite{kaplan2002statistics} and found in microwave transport experiments \cite{barkhofen2013}.

\section{Branch Formation and Stability in Disorder Potentials}\label{sec:mechDis}

\begin{figure}
\centering
\includegraphics[width=0.95\columnwidth]{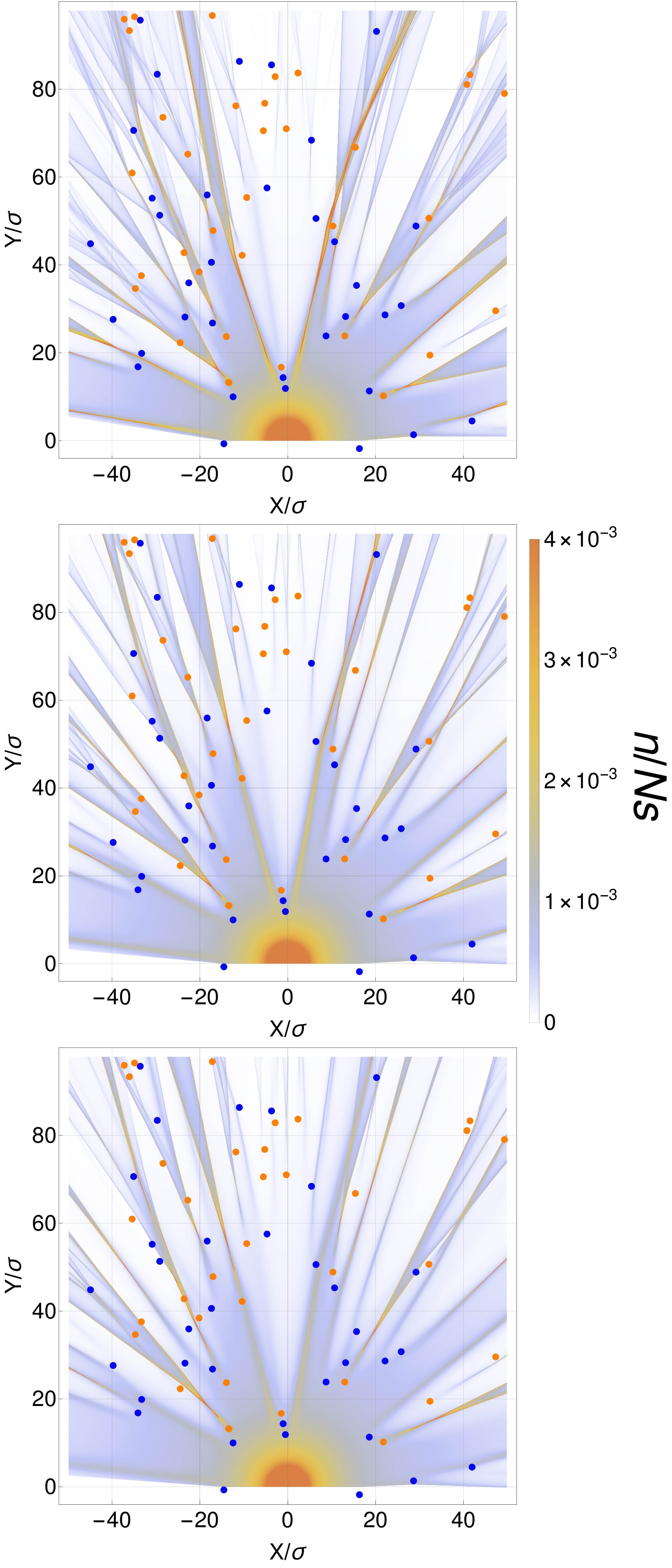}
\caption{Density of classical electron trajectories as a result of scattering from 70 randomly-placed, localized Gaussian bumps, half of them repulsive (blue dots), and half of them attractive (orange dots). The numerical simulation involves $100,000$ trajectories distributed uniformly over an angular range of $\pi$ radians, thus $N \approx 31,831$ trajectories per radian. From top to bottom, the energies are 5, 8, and 10 $V_{0}$. The choice of plotting parameters is the same as in Fig.\ \ref{fig:zB}.}
\label{fig:mBS}
\end{figure}
\begin{figure}
\centering
\includegraphics[width=0.95\columnwidth]{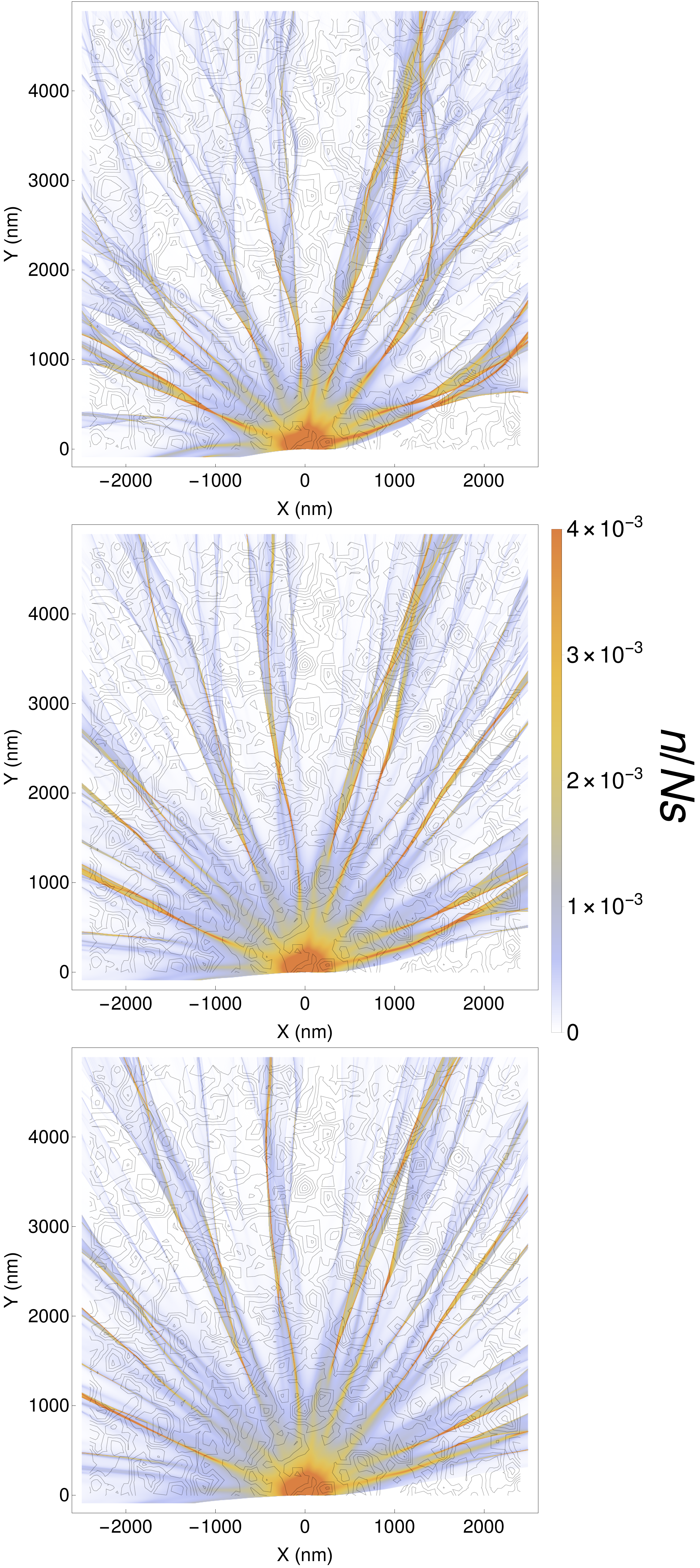}
\caption{Density of classical electron trajectories as a result of scattering from a smooth disorder potential, the details of which can be found in Appendix \ref{app:disGen}. The black curves indicate contour lines of the potential. The choice of parameters is the same as in Fig.\ \ref{fig:mBS}. The maximum amplitude of the disorder potential is approximately 1~meV. The characteristic width $\sigma$ in this case is roughly set by the distance to the doping layer in our simulation of the realistic disorder, which in this case is taken to be 70~nm.}
\label{fig:SRD}
\end{figure}
In the previous sections we saw that a generic localized feature leads to the formation of branches in the SGM response which are robust with respect to changes of the electron energy. As discussed at the end of Sec.\ \ref{subsec:localmax}, when going beyond the case of our toy model, by considering other scattering centers, new branches could be generated from the residual flux of the first-appearing branch. In order to study how the mechanisms behind branch formation and energy stability carry over to the experimentally investigated 2DEG, we present two generalizations towards a realistic description of a smooth disorder potential. Firstly, we consider a disorder potential modeled by a collection of several randomly placed repulsive and attractive Gaussian features, all with the same amplitude and width as before. Secondly, we provide results for the more realistic modeling of the disorder potential described in Appendix \ref{app:disGen}.

Figure \ref{fig:mBS} displays the trajectory density resulting from a point-like QPC emitting a uniform angular distribution of classical trajectories, for the model of several local features, with the energy of the electrons increasing from top to bottom. This simple model allows us to make the transition from the one-feature phenomena to a genuine disorder potential. The closest features verifying the criterion \eqref{eq:impulse11} give rise to first-generation branches. These branches might be affected by other features downstream and, as explained before, new branches might appear from the residual flux. The combined effects of repulsive and attractive features result in a spatial pattern which is similar to that of more refined models (i.e. Fig.\ \ref{fig:SRD}) or the experimentally obtained SGM scans. At this point it is important to remark that the quantitative description of the branching phenomena is limited by the lack of a satisfactory definition of what it means to be in a branch (as opposed to not being in one), as well as by the obvious fact that not all the contributing trajectories belong to branches \cite{heller2003branching}. For instance, we remark that the threshold used to represent the trajectory density considerably affects the characteristics of the corresponding mapping. This is why the quantitative analysis aimed at in this work, concerning the branch formation and the associated energy stability, becomes particularly relevant.

The robustness associated with the individual repulsive and attractive features results in an important energy stability, since the branching pattern remain largely fixed over the large range of chosen energies. An even closer resemblance with the experimentally obtained branching pattern can be achieved when modeling the disorder potential in a more realistic fashion (Fig.\ \ref{fig:SRD}). This last model is intended to represent the disorder created by a collection of impurities distributed randomly in the dopant layer at a distance of 70~nm from the 2DEG (the details of this disorder potential and its generation can be found in Appendix \ref{app:disGen}). The contours in the background of the plots highlight the features of the disorder potential. We observe the same qualitative features of branching as in our previous simpler model, including the stability of many of the branches with respect to energy. Notice, however, that many of the branches have variable width under energy changes, becoming more focused in the forward direction as the energy is raised, a feature consistent with the behavior of branches formed by scattering from a localized attractive feature, while other branches remained almost unchanged in width, consistent with the behavior of scattering from a repulsive potential. This emphasizes the need to consider both cases in order to understand the phenomena of branch formation and energy stability. 

It is important to note that it is not always possible to identify the formation of a branch in Fig.\ \ref{fig:SRD} with a specific, isolated minimum or maximum of the potential. Indeed, a realistic model of disorder will necessarily involve features significantly more complicated than this. As explained in Sec.\ \ref{sec:mech}, however, the key feature of our toy model which allows for branch formation regards the nature in which the geometry of the edge of the potential varies with respect to other relevant parameters of the model, a mechanism which is not specific to isolated peaks.

The criterion \eqref{eq:impulse11} put forward for individual features remains useful in order to estimate the distance for the appearance of the first branches in the case of more realistic models of the disorder potential.
For typical experiments, the distance to the doping layer determines the scale of variation of the smooth potential \cite{jura2007unexpected} (and thus a minimum ``bump width''), and the energy of the scattered electrons is an order of magnitude larger than the amplitude of the fluctuations of the disorder potential. These estimations, when combined with the condition \eqref{eq:impulse11}, predict that the scale over which branches first appear should be about an order of magnitude larger than the distance to the doping layer, which is consistent both with experiment \cite{braem2018stable,topinka2003imaging,heller2003branching,jura2007unexpected,topinka2001coherent}, and our numerical simulations of smooth disorder.

As the energy of the electrons is lowered, the threshold distance for branch formation will also be reduced. As local features closer to the QPC begin to form new branched structures, some of the flux of electron trajectories will be focused into these branches, before reaching bumps in the potential which are further from the QPC. Thus, the relative intensity of some branches may change slightly with energy, while the overall branching structure remains static in space. This is again observed in experiment \cite{braem2018stable}, as well as our own numerical simulations, as displayed in Figs.\ \ref{fig:mBS} and \ref{fig:SRD}. Each scattering center which is sufficiently far from the QPC to meet the branching criteria outlined previously, yet close enough to receive a sufficiently large portion of the flux from the QPC, will result in a narrowly focused branch of classical trajectories.

The pertinence of the latter model of smooth disorder can also be appreciated since, as shown in Appendix \ref{app:SS}, it yields results which are consistent with experiments \cite{jura2007unexpected} and previous theoretical work \cite{jura2007unexpected,liu2013stability} indicating that the branching pattern at large distances is stable against a lateral shift of the QPC in physical space.

\section{Concluding Remarks}
\label{sec:con}

In this work we have provided an explanation for the robust stability of branches in the scanning gate response of two-dimensional electron gases with smooth disorder, with respect to a change in the Fermi energy observed in Ref.\ \cite{braem2018stable}. We have done so by first invoking a toy model for the formation and the energy stability of these branches, which we have argued is sufficient to capture all of the observed features of branching in more refined models and in the experiments. We have found that the stability of these branches is extremely generic, and does not rely on the detailed shape of the disorder potential, but only upon the assumptions of weak scattering, weak electron interactions, and the hypothesis that the SGM response can be interpreted as being proportional to the local density of classical trajectories. 

The quantitative criteria for branch formation and energy stability, found for a toy model of a single localized feature, provide valuable insight for the case of more elaborated descriptions of the smooth disorder present in the 2DEG.
Our findings could have applications in probing the nature of the disorder potential in setups of two-dimensional electron gases other than that of GaAs heterostructures, for example, those created in samples of bi-layer graphene \cite{doi:10.1021/acs.nanolett.7b04666,PhysRevLett.121.257702}.
Moreover, even if we have studied the specific case of disordered electron gases, our methods are quite general, and should be equally applicable to the wide range of other physical systems mentioned in the introductory remarks. It is our hope that our work may aid towards a more detailed understanding of the precise microscopic processes which give rise to branched flow and its prominent robustness.

\begin{acknowledgments}

We thank Beat Braem, Klaus Ensslin, Carolin Gold, Thomas Ihn, Steven Tomsovic, and Guillaume Weick for helpful discussions. This work was funded by the French National Research Agency (ANR) through the Programme d'Investissement d'Avenir under contract ANR-11-LABX-0058\_NIE within the Investissement d’Avenir program ANR-10-IDEX-0002-02, and through the Grant  ANR-14CE36-0007-01 (project SGM-Bal).

\end{acknowledgments}


\appendix

\section{Generation of smooth disorder and numerical simulation of trajectory density}
\label{app:disGen}

We describe in this Appendix the generation of the smooth disorder potential which appears in Sec.\ \ref{sec:mechDis}, following the lines of Ref.\ \cite{ihn2010semiconductor}, as well as the numerical simulation used to compute the classical trajectories in both this disorder potential, and also the toy model presented in Sec.\ \ref{sec:mech}.

We assume that the disorder is caused by randomly distributed singly-ionized dopants in the doping plane of the semiconductor heterostructure used to generate the 2DEG \cite{ihn2010semiconductor,PhysRevB.41.7929}. We will assume the 2DEG is a square with side length $L$, and a total number of dopants $M$, for an average dopant density of 
\begin{equation}
n_\mathrm{d} = M / L^{2}.
\end{equation}

Under such conditions, the screened potential in the plane of the electron gas can be shown \cite{ihn2010semiconductor} to take the form
\begin{equation}
\begin{split}
& V\left ( \vec{r} \right ) = \\ &-2 \frac{\left (\Delta q \right )^{2}}{\pi}E_{\text{Ryd}}^{*}a_{\text{B}}^{*} \sum_{q_{j} > 0}  \frac{e^{-q_{j}s}}{q_{j}+q_{\text{TF}}}R_{j} \cos \left ( \vec{q}_{j} \cdot \vec{r}+\phi_{j} \right ).
\end{split}
\label{eq:screenPot}
\end{equation}
Here $E_{\text{Ryd}}^{*}$ is the effective Rydberg energy, $a_{\text{B}}^{*}$ the effective Bohr radius, and the distance between the electron gas and the doping layer is given by $s$. The vectors $\vec{q}_{j}$ live on a discrete lattice in Fourier space with lattice spacing $\Delta q = 2 \pi / L$, and $q_{\text{TF}}=2  / a_{\text{B}}^{*}$. The terms $R_{j}$ and $\phi_{j}$ form a set of random amplitudes and phases that define a complex variable $\widetilde{C}\left ( \vec{q}_{j} \right ) \equiv R_{j}e^{i \phi_{j}}$ associated to each Fourier lattice vector $\vec{q}_{j}$. This complex variable is equal to the Fourier transform of $C_{2}$, which is the projection of the fluctuating part of the charge distribution into the two-dimensional plane of the doping layer,
\begin{equation}
C_{3} \left( \vec{\rho} \right) \equiv \left[ \sum_{i=1}^{M} \delta \left( \vec{r} - \vec{r}_{i}  \right) - n_\mathrm{d} \right] \delta \left( z-s \right) \equiv C_{2} \left( \vec{r} \right)\delta \left( z-s \right),
\end{equation}
where $\vec{r}$ is a point in a two-dimensional plane parallel to the electron gas and the doping layer, $z$ is the coordinate direction perpendicular to this plane, with $z=0$ at the location of the electron gas, $\vec{\rho} \equiv \left ( \vec{r},z \right )$ is a point in the three-dimensional heterostructure, and the vectors $\vec{r}_{i}$ are the locations of the randomly distributed dopants.

Due to the exponential term in the summation in Eq.\ \eqref{eq:screenPot}, large Fourier modes do not contribute substantially, which allows for a significant truncation of the sum. Notice that this effectively suppresses fluctuations of the potential on length scales shorter than the spacing $s$.

Due to the large number of dopants which are typically present in a realistic heterostructure, it would not be computationally tractable to randomly select a collection of dopant positions and compute all of their contributions to $\widetilde{C}$. Thus, instead of selecting a collection of random dopants, we directly study the statistical properties of $\widetilde{C}$. From the definition of $\widetilde{C}$, along with the fact that the original dopants are uniformly distributed and there is a macroscopically large number of them, it is a straight-forward exercise \cite{ihn2010semiconductor} to invoke the central limit theorem and find that the real and imaginary parts of $\widetilde{C}$ are both normally distributed, with mean zero and variance
\begin{equation}
\sigma ^{2} = M / 2.
\end{equation}
In our computation of the disorder potential, we therefore draw a random distribution of real and imaginary terms (which we use to compute the phase angles and amplitudes), for sufficiently small lattice momentum.

\begin{figure}[t!p]
\centering
\includegraphics[width=85mm]{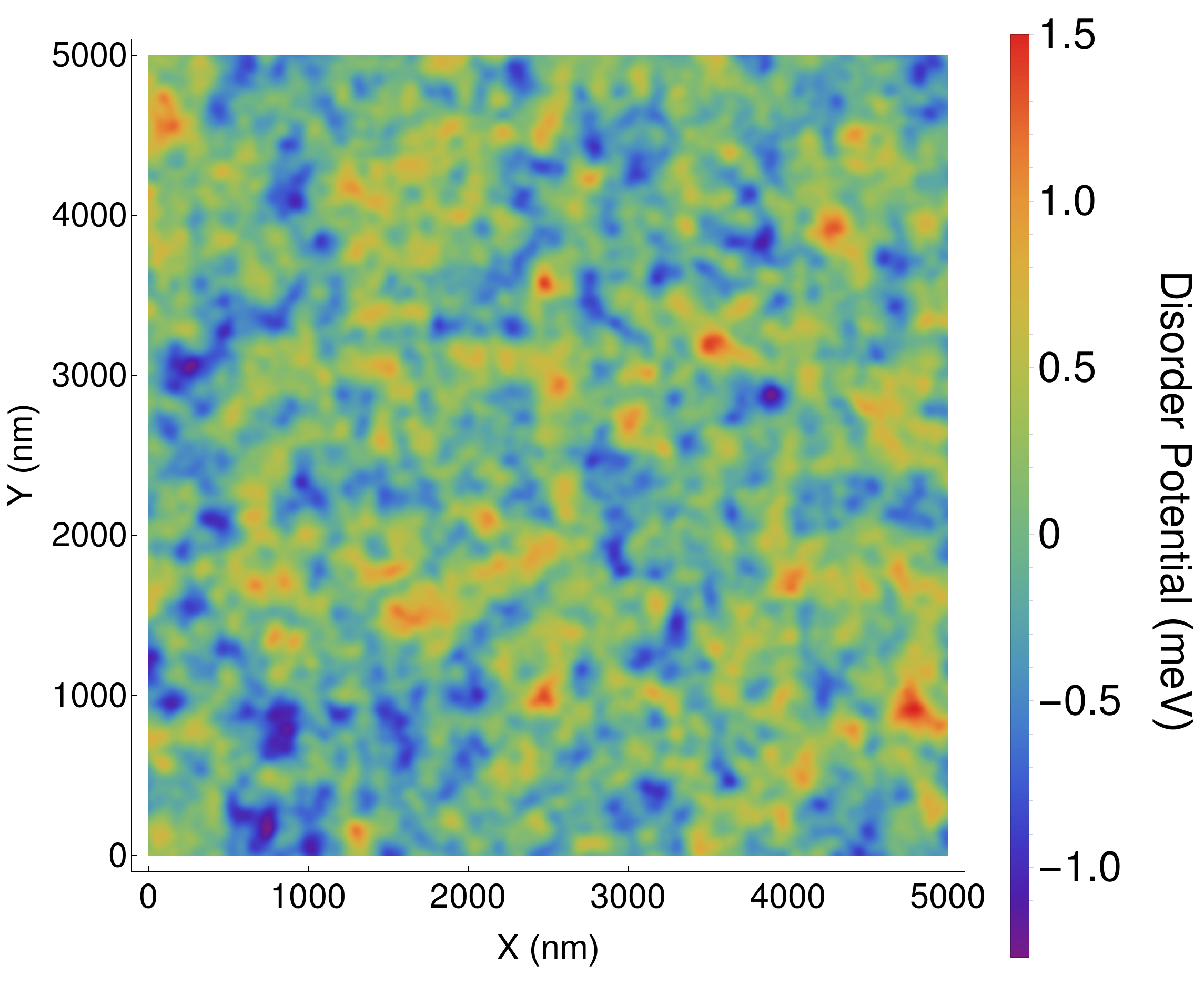}
\caption{Intensity of the smooth disorder potential used in the simulation of our disordered 2DEG according to the color scale indicated at the right (in meV).}
\label{fig:dis}
\end{figure}

Figure \ref{fig:dis} shows a plot of the disorder potential which results from these calculations. It is generated as a result of choosing $M=$150,000, $L = 10 ~\mu$m, and $s = 70$~nm, along with $E_{\text{Ryd}}^{*} = 5.76 ~$meV and $a_{\text{B}}^{*}  = 10$~nm, all chosen to match experimentally realistic values. The resulting potential has a maximum amplitude of approximately 1.53 meV, an RMS amplitude of approximately 0.37 meV, and a mean of zero. In our trajectory simulations, we use a square patch within this disorder, with dimensions 5 micrometers by 5 micrometers.

For the simulation of classical trajectories, we use a standard fourth-order Runge-Kutta iterator. For the toy model presented in Sec.\ \ref{sec:mech}, we simply evaluate the potential at each Runge-Kutta iteration. However, due to the large number of cosine terms which must be computed for each location in the disorder sample, it would be computationally infeasible to evaluate the value of the disorder potential at every point along every trajectory that we simulate. Instead, we perform only one computation of the disorder potential, along with its first, second, and third order mixed derivatives, on a lattice with a spacing of one nanometer. Since the higher order derivatives of the disorder potential contain the same sine and cosine terms as the original sum and its first derivative, we can obtain these higher order derivatives at effectively no extra computational cost. These values are then saved and reused for each trajectory simulation. During the simulation, the value of the disorder at any point along a trajectory is computed by finding the closest lattice point, and performing a Taylor series approximation. Since the derivatives of the disorder potential have already been computed at each lattice point, this amounts to a simple algebraic sum, as opposed to a full Fourier series. Our benchmarking indicates that this technique allows for the trajectory simulations to be numerically tractable, with an error which is essentially negligible compared with the full computation of the potential for each point along the trajectory.

For both models of disorder, the density of trajectories is computed by counting the number of trajectories which pass within a radius of $r = s / 2$ of a given lattice site on a grid of spacing $s$. This number is associated with the density of trajectories at that lattice point.

\section{Equivalence between the branch formation criterion of Sec.\ \ref{sec:crit} and the condition for formation of a caustic}
\label{sec:caustEq}

Here we elaborate on the claim of Sec.\ \ref{sec:crit} that the mathematical criterion we have found for a divergence in the density of classical trajectories is approximately equivalent to the condition for caustic formation. 

The condition for the formation of a caustic is given by Eq.\ \eqref{eq:caustic} of the main text, where $q \left ( t \right )$ is the position of some scattered electron at fixed time $t$, and $p_\mathrm{i}$ is its initial momentum. Expressing the electron trajectories in polar coordinates $\left ( \vec{r},t \right )$, which are themselves functions of the initial angle $\alpha_\mathrm{i}$, the initial momentum $p = \sqrt{2mE}$, and the time $t$, we write
\begin{equation}
q_{x} \left ( t \right ) = r \left ( t \right ) \sin \left ( \alpha \left ( t \right ) \right )~;~q_{y} \left ( t \right ) = r \left ( t \right ) \cos \left ( \alpha \left ( t \right ) \right ).
\end{equation}
Equation \ref{eq:caustic} becomes, after some straight-forward algebra,
\begin{equation}
\frac{\partial \alpha}{\partial p}\frac{\partial r}{\partial \alpha_\mathrm{i}} - \frac{\partial \alpha}{\partial \alpha_\mathrm{i}}\frac{\partial r}{\partial p} = 0.
\end{equation}
At sufficiently large times at which we can associate $\alpha \left ( t \right )$ with $\alpha_\mathrm{f}$, this becomes
\begin{equation}
\frac{\partial \alpha_\mathrm{f}}{\partial p}\frac{\partial r}{\partial \alpha_\mathrm{i}} - \frac{\partial \alpha_\mathrm{f}}{\partial \alpha_\mathrm{i}}\frac{\partial r}{\partial p} = 0.
\end{equation}

For weak scattering, the radius at a fixed large time is only weakly dependent on initial scattering angle, and so satisfaction of the above equation is roughly tantamount to
\begin{equation}
\frac{\partial \alpha_\mathrm{f}}{\partial \alpha_\mathrm{i}} = 0,
\end{equation}
as claimed in the main text.

\section{Computation of the scattering function and the critical energy and distance relationship}
\label{app:scatterApprox}

Here we derive the scattering function and its form in the impulse approximation, and also briefly elaborate on the relationship we have displayed in the main text regarding the critical energy and distance relationship exhibited in Eq.\ \eqref{eq:impulse11}.

As a result of energy and angular momentum conservation in our system, we can write
\begin{equation}
E = \frac{1}{2}m\dot{r}^{2} +  V_{\text{eff}} \left ( r \right )~;~\frac{d\theta}{dt} = l / mr^{2},
\end{equation}
where $r$ is the distance from the center of the localized scattering feature, and $\theta$ is the polar angle defined with respect to the axis between the QPC and the scattering center. The electron mass is given by $m$, and $l$ is the angular momentum. The effective potential is given
\begin{equation}
V_{\text{eff}} \left ( r \right ) = \frac{1}{2}\frac{l^{2}}{mr^{2}}  + V \left ( r \right ).
\end{equation}
Using the last two equations to eliminate time from our problem, we find
\begin{equation}
d \theta =  \pm \frac{l}{\sqrt{2m}}\frac{dr}{r^{2}\sqrt{E-V_{\text{eff}}\left ( r \right )}}.
\end{equation}
Integrating this equation from infinity to the radius of closest approach $r_{*}$, and then out to infinity again, we find equations \ref{eq:scattAngMain} through \ref{eq:paramDef1} of the main text in Sec.\ \ref{sec:enStabGen}. Equation \ref{eq:paramDef1} comes as a result of the requirement that the radius of closest approach is a turning point of the effective one-dimensional potential.

In order to obtain the approximate form of this expression in the limit of weak scattering, we can take two equivalent approaches. First, it is possible to simply Taylor expand the square root in Eq.\ \eqref{eq:intDefMain} in the limit of small $V/E$. Along with this, we assume an expansion of the parameter $\lambda$ in powers of $V/E$, and take $\lambda = 1$ to lowest order. Using these approximations, and performing an integration by parts on the integral, we eventually arrive at the approximate form (\ref{eq:fDef}) given in the main text.  

Alternatively, it is possible to obtain this approximate form from first principles, using an impulse approximation, which we briefly outline here. For simplicity, we redefine our coordinate axes slightly, so that the momentum along the direction of propagation is taken to be $p_{y}$. The force acting on the particle in the $x$-direction, transverse to the propagation, is $F_{x}$, and the momentum $p_{x}$ gained by the particle  after scattering from the potential is given by
\begin{equation}
p_{x} = \int_{-\infty}^{\infty} dt F_{x} \left ( t \right ) = - \int_{-\infty}^{\infty} dt \left . \frac{\partial V}{\partial x} \right |_{x=b,y=tp_{y}/m},
\label{eq:impulse1}
\end{equation}
where $V$ is the potential. Using $x=b$ and $y=tp_{y}/m$, and assuming that the potential is radial, $V \left ( x, y \right ) \equiv V \left ( r \right )$, we find
\begin{eqnarray}
p_{x} &=& \int_{-\infty}^{\infty} dt\frac{x}{r}\frac{\partial V}{\partial r}\nonumber\\
 &=& -\frac{mb}{p_{y}}\int_{-\infty}^{\infty} dy \frac{1}{\sqrt{b^{2}+y^{2}}}\left . \frac{\partial V}{\partial r} \right |_{r = \sqrt{b^{2}+y^{2}}}.
\label{eq:impulse2}
\end{eqnarray}
With some additional rearrangement, and the approximation
\begin{equation}
b =a \sin \alpha_\mathrm{i} \approx a \alpha_\mathrm{i},
\label{eq:impulse4}
\end{equation}
this result can be stated as
\begin{equation}
p_{x} = -\frac{2m}{p_{y}}\int_{a \alpha_\mathrm{i}}^{\infty}\frac{dr}{\sqrt{\left ( r/a \alpha_\mathrm{i} \right )^{2}-1}}\frac{\partial  V}{\partial r}
\label{eq:impulse5}
\end{equation}

\begin{figure}
\centering
\includegraphics[width=0.9\columnwidth]{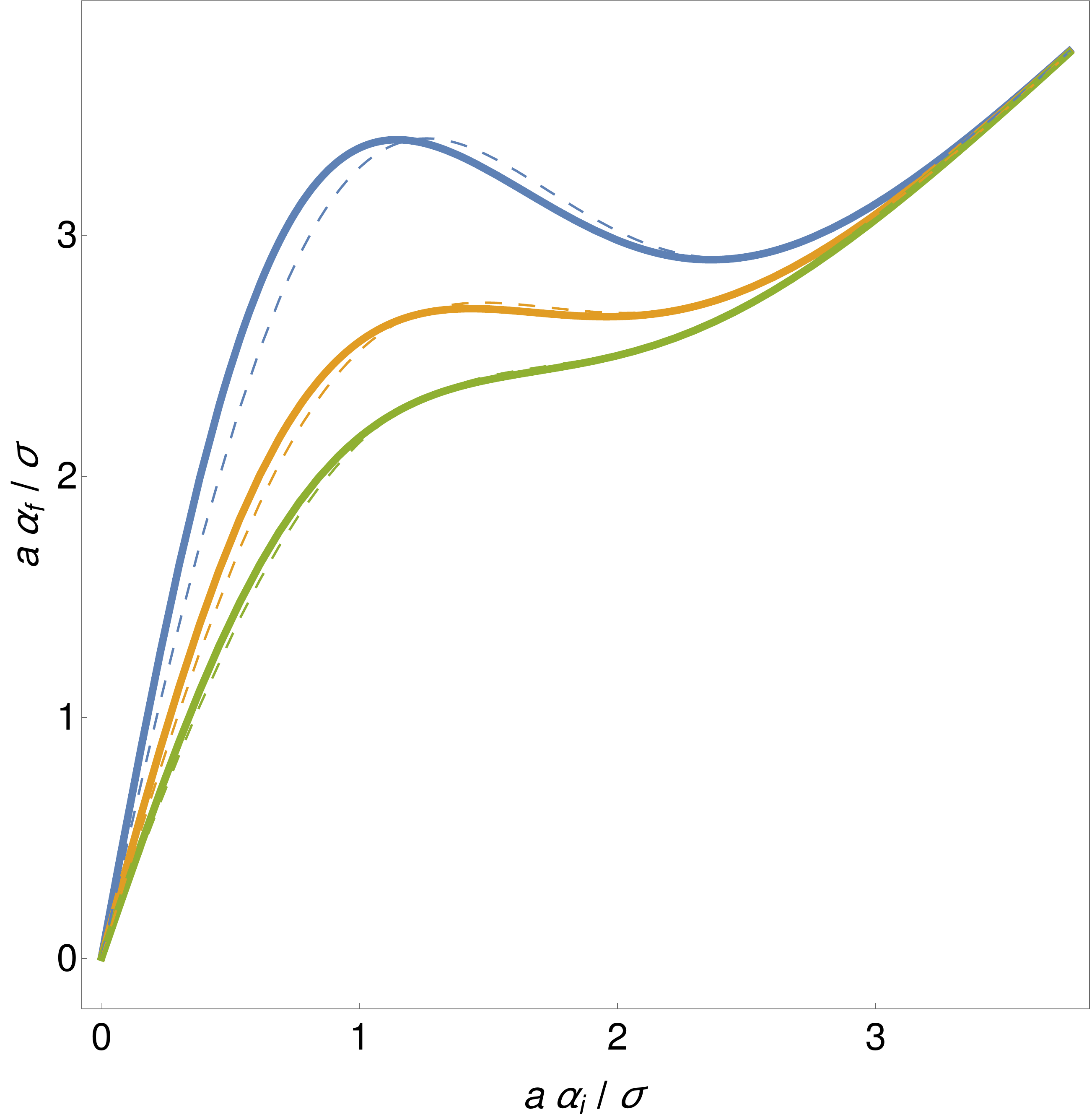}
\caption{A comparison of the exact and approximate scattering functions, for the case of a repulsive Gaussian bump of width $\sigma$, placed a distance $a=15\sigma$ from the QPC. The solid lines represent the exact scattering function, while the dashed lines represent the approximate scattering function. The energies, from top to bottom, are 5, 7.5, and 10 $V_{0}$.}
\label{fig:sA}
\end{figure}
With this result for the change in transverse momentum, we can find the net deflection as
\begin{equation}
p_{x} / p_{y} = \tan \delta \alpha \approx \delta \alpha ,
\label{eq:impulse6}
\end{equation}
where we have assumed that the deflection is small. Thus, in this approximation,
\begin{equation}
\alpha_\mathrm{f} = \alpha_\mathrm{i} + \delta \alpha \approx \alpha_\mathrm{i}  -\frac{2m}{\sigma p_{y}^{2}}\int_{a \alpha_\mathrm{i}}^{\infty}\frac{dr}{\sqrt{\left ( r/a \alpha_\mathrm{i} \right )^{2}-1}}\frac{\partial  V}{\partial \left ( r / \sigma \right )},
\label{eq:impulse7}
\end{equation}
or,
\begin{equation}
\alpha_\mathrm{f}  \approx \alpha_\mathrm{i}  -\frac{a \alpha_\mathrm{i}}{\sigma E}\int_{1}^{\infty}\frac{ds}{\sqrt{s^{2}-1}}V ' \left ( a \alpha_\mathrm{i} s \right ).
\label{eq:impulse8}
\end{equation}
In Fig.\ \ref{fig:sA}, which compares the full scattering expression to the approximate one, we see that this is indeed a very accurate approximation for the parameter regime we are interested in.  

This approximate integral can now be solved exactly for some special choices of the model potential. In particular, we will focus on two example cases, a Lorentzian potential hill,
\begin{equation}
V \left ( r \right ) = V_{0}\frac{\sigma^{2}}{r^{2}+\sigma^{2}},
\label{eq:impulse9}
\end{equation}
and a Gaussian hill,
\begin{equation}
V \left ( r \right ) = V_{0} \exp \left [ -r^{2} / 2 \sigma^{2} \right ].
\label{eq:impulse10}
\end{equation}
For the Lorentzian, we find
\begin{equation}
\alpha_\mathrm{f}^{L} \approx \alpha_\mathrm{i} + \frac{\pi}{2}\frac{V_{0}}{E}\frac{a \alpha_\mathrm{i}\sigma^{2}}{\left [ \left ( a \alpha_\mathrm{i} \right )^{2} + \sigma^{2} \right ]^{3/2}},
\end{equation}
while for the Gaussian, we find
\begin{equation}
\alpha_\mathrm{f}^{G} \approx \alpha_\mathrm{i} + \sqrt{\frac{\pi}{2}}\frac{V_{0}}{E}\frac{a \alpha_\mathrm{i}}{\sigma} \exp \left [ - \left ( a\alpha_\mathrm{i} \right )^{2} / 2 \sigma^{2} \right ].
\end{equation}
The condition for a zero derivative then becomes, for the case of the Lorentzian,
\begin{equation}
1 + \frac{\pi}{2}\frac{V_{0}}{E}\frac{a \sigma^{4}}{\left [ \left ( a \alpha_\mathrm{i} \right )^{2} + \sigma^{2} \right ]^{5/2}} \left ( 1 - 2 \frac{a^{2}}{\sigma^{2}} \alpha_\mathrm{i}^{2}\right ) = 0,
\end{equation}
and for the case of the Gaussian,
\begin{equation}
1 + \sqrt{\frac{\pi}{2}}\frac{V_{0}}{E}\frac{a }{\sigma} \exp \left [ - \left ( a\alpha_\mathrm{i} \right )^{2} / 2 \sigma^{2} \right ] \left ( 1 - \frac{a^{2}}{\sigma^{2}} \alpha_\mathrm{i}^{2}\right ) = 0.
\end{equation}
In both cases, the prefactor on the second term in the expression must be of order one in order for this condition to be satisfied. Examination of these terms reveals that this will be the case so long as Eq.\ \eqref{eq:impulse11} is fulfilled.

\section{Stability of the branching pattern with respect to a physical shift of the QPC}
\label{app:SS}

We address in this appendix a feature of branch formation in realistic models of disorder, not treated in the main text, which is nonetheless consistent with our toy model. Experimental and numerical studies of electron flow in disordered potentials resulted in branching patterns at large distances that are stable against a lateral shift of the QPC in physical space \cite{jura2007unexpected}. Previous authors have identified the finite width of the QPC as being necessary for explaining this observed behavior, in particular, requiring a width which is roughly the same order of magnitude as the size of the shift \cite{jura2007unexpected,liu2013stability}. This phenomenon, thoroughly investigated by these authors, constitutes a good test case for our proposed model of branch formation. 

The stability with respect to the lateral displacement of the QPC observed in Ref.\ \cite{jura2007unexpected} could be considered, in first sight, surprising since the chaotic nature of the underlying classical electron dynamics goes together with an extreme sensitivity to the initial conditions. The coherent overlap between two wave-packets representing the evolution associated with different Hamiltonian was found to become sizeable at some distance from the QPC and remain significant even very far away \cite{liu2013stability}. And the same kind of argument was proposed to be applicable in the classical case.  

\begin{figure}
\centering
\includegraphics[width=85mm]{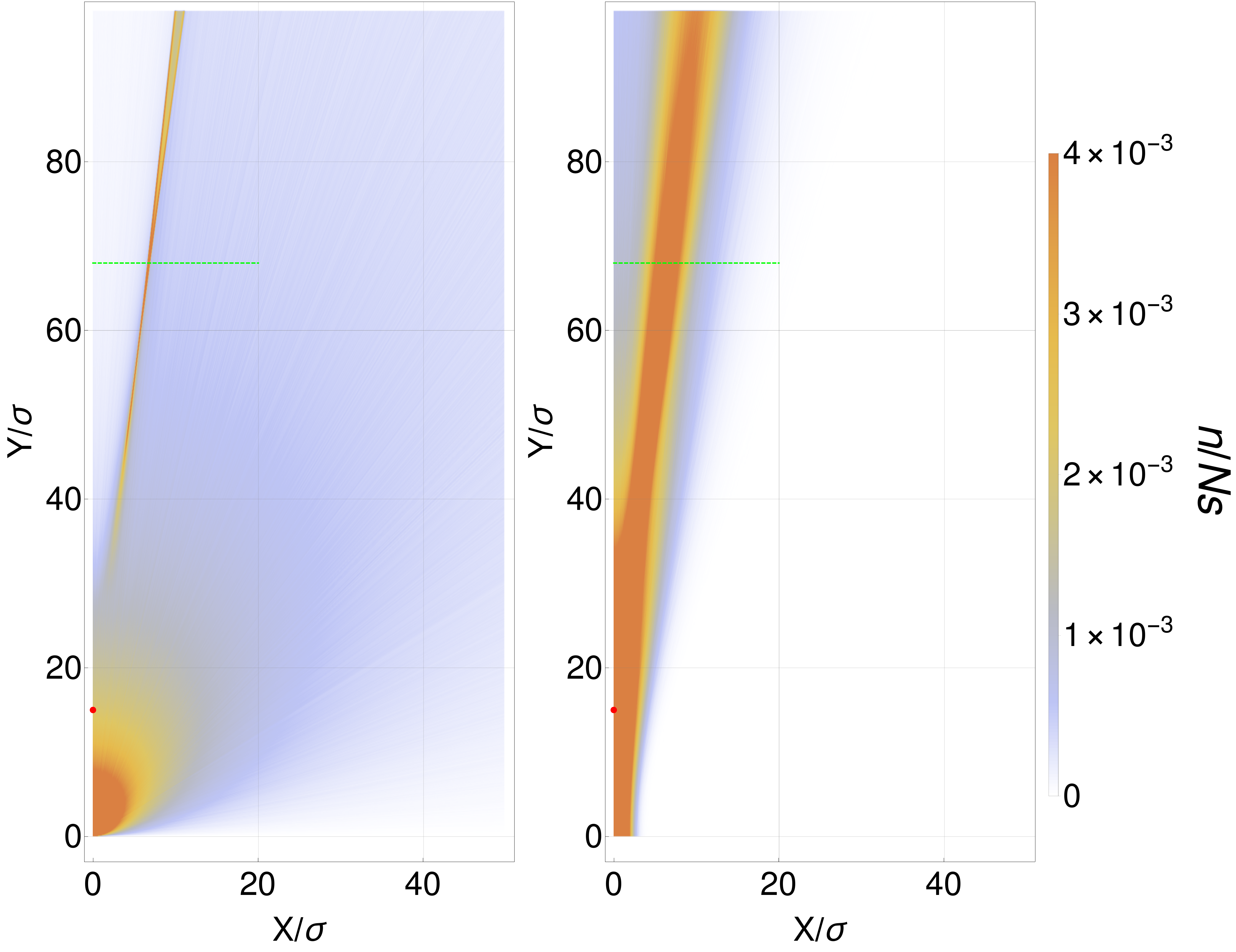} 
\caption{Density of classical electron trajectories as a result of scattering from a localized, repulsive Gaussian bump (red dot). The distance between the center of the bump and the QPC is taken to be $a = 15 \sigma$, with the center of the QPC taken to be at the origin. The QPC is modeled as having a finite width of 0.02$\sigma$ (left) and $\sigma$ (right). The numerical simulation involves 1,000,000 trajectories distributed non-uniformly over a range of $\pi$ radians (which results in an average $N \approx 318,310$ trajectories per radian), all with an energy of 7.5 $V_{0}$; the details of the probability distribution of electron initial conditions that we use can be found in the main text of the appendix. The choice of plotting parameters is the same as in Fig.\ \ref{fig:zBN}}.
\label{fig:FW}
\end{figure}
To study the more physically realistic case of a QPC with a finite opening width, we use an approach previously outlined by Liu and Heller \cite{liu2013stability}. For a QPC with harmonic confinement, the Hamiltonian is given
\begin{equation}
H_{0} = \frac{p^{2}}{2m}+\frac{1}{2}m [\omega(y)]^{2}x^{2},
\end{equation}
where $\omega(y)$ is a slowly varying function of $y$, which decreases as the QPC opens. For electrons with a given energy, the Wigner quasiprobability distribution associated with the scattering eigenstates yields
\begin{equation}
P \left ( x, p_{x} \right ) = \frac{1}{\pi \sigma_{p_{x}}\sigma_{x}}\exp \left [ - \left ( p_{x}^{2}/\sigma^{2}_{p_{x}} \right )- \left ( x^{2} / \sigma^{2}_{x} \right ) \right ],
\end{equation}
where $\sigma_{p_{x}}$ and $\sigma_{x}$ are related by 
\begin{equation}
\sigma_{p_{x}} = \hbar / \sigma_{x},
\end{equation}
and can be determined from the bare parameters of the Hamiltonian, if desired.

\begin{figure}
\centering
\includegraphics[width=85mm]{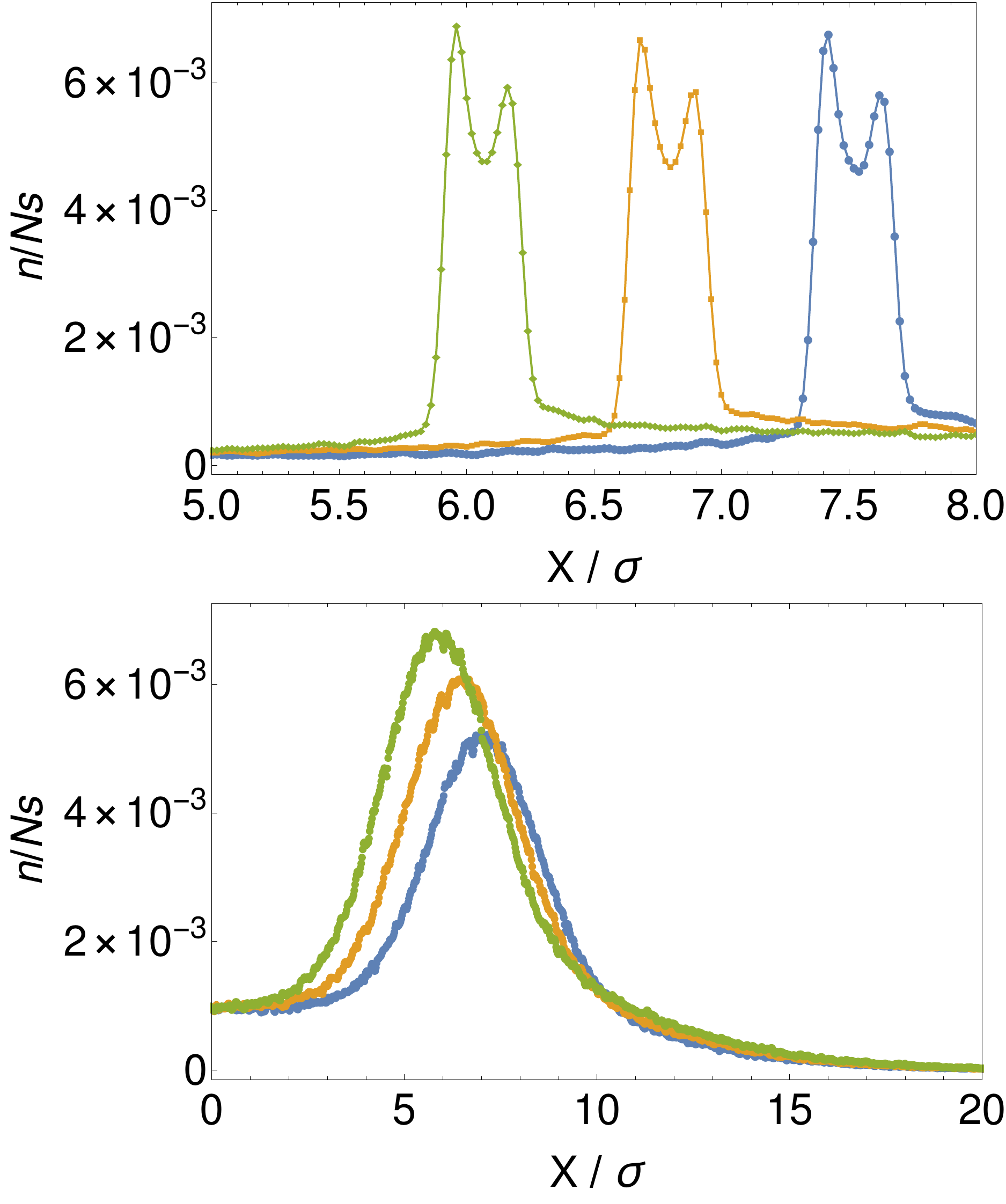} 
\caption{Density of trajectories across the horizontal cut displayed in Fig.\ \ref{fig:FW}, located at a vertical position of $y = 68 \sigma$. Each panel displays the density after having shifted the QPCs in Fig.\ \ref{fig:FW} either 0.5$\sigma$ to the left (blue dots) or to the right (green dots), as well as the case of no shift at all (orange dots). The top panel displays the results for the QPC with a width of 0.02$\sigma$, while the bottom panel displays the results for the QPC with a width of $\sigma$. The energy of the electrons in all cases is 7.5~$V_{0}$.}
\label{fig:FWcomp}
\end{figure}
Here, we will choose to set $\sigma_{x}$ to values of 0.02 and 1 times the width of the Gaussian bump. Following Ref.\ \cite{liu2013stability}, we perform our numerical simulation by randomly selecting initial $x$ and $p_{x}$ according to the probability distribution above. For a given electron energy $E$, we eliminate any randomly selected $p_{x}$ which result in a total electron energy larger than $E$, and we boost all other trajectories in $p_{y}$ such that the total electron energy is $E$. The vertical starting position $y$ of all electron trajectories is taken to be the same. After the generation of such a random electron initial condition, we propagate the corresponding trajectory classically.

Our numerical results indicate that the stability with respect to a lateral shift of the QPC is present in our toy model. Figure \ref{fig:FW} displays the trajectory density for a Gaussian bump, with the QPC being modeled to have two different widths. We note that the overall branching structure is qualitatively similar, with the branch becoming wider for a QPC with larger width. Figure \ref{fig:FWcomp} displays the density of trajectories across the horizontal cuts indicated in Fig.\ \ref{fig:FW}, when the QPC is shifted to the left and right by an amount comparable to the width of the wider QPC. In fact, the change in branch position is essentially due to a tilt in the axis connecting the QPC center with the potential hill, which is small when the shift is small as compared to the QPC-hill distance. For a more realistic disorder potential, a similar effect should occur with respect to the potential features that are responsible for branch formation. While the location of the branch is relatively unstable for the case of a narrow QPC, it becomes broader, yet significantly more stable, for the case of a QPC whose width is chosen to be the same size as the shift in physical space. This is consistent with the results of previous authors \cite{jura2007unexpected,liu2013stability} studying more realistic models of disorder, and provides further evidence that our proposed mechanism is capable of capturing the correct physics of branch formation.

We have also found that the stability with respect to a QPC shift is consistent with the presence of a localized attractive feature. We note that, since the effect of a finite width QPC is to broaden the branches, it is possible that some of the lack of stability in the width of the branches formed by an attractive feature could be less noticeable if the branches are washed out over a distance scale comparable to the separation between the two branches on either side of the attractive feature. However, since this mechanism would rely on the detailed nature of the QPC, we will not investigate it here, as we are interested in a more generic understanding of branch stability. 


\bibliography{paper_references}

\end{document}